\documentclass[aps,twocolumn,superscriptaddress,showpacs,showkeys]{revtex4}
\usepackage{graphics,graphicx,dcolumn,bm,fleqn,epic,eepic,float,epsfig}
\usepackage{amssymb,amsmath,multirow,rotate,color,float}
\usepackage{epstopdf}
\usepackage{times}
\usepackage{color}
\usepackage{soul}                    
\definecolor{red}{rgb}{1,0,0}
\definecolor{green}{rgb}{0,1,0}
\definecolor{blue}{rgb}{0,0,1}

\newcommand{\f}{\mathbf}

\begin{document}

\title{Uncovering wind turbine properties through two-dimensional
stochastic modeling of wind dynamics}

\author{Frank Raischel}
\affiliation{Center for Theoretical and Computational Physics,
        University of Lisbon, Av.~Prof.~Gama Pinto 2,
        1649-003 Lisbon, Portugal}
\affiliation{Center for Geophysics, IDL, University of Lisbon
         1749-016 Lisboa, Portugal}
\author{Teresa Scholz}
\affiliation{Energy Systems Modeling and Optimization Unit,
        National Laboratory for Energy and Geology (LNEG),
        Estrada do Pa\c{c}o do Lumiar 22, 1649-038 Lisbon, Portugal}
\affiliation{Departamento de F\'{\i}sica, Faculdade de Ci\^encias 
             da Universidade de Lisboa, 1649-003 Lisboa, Portugal} 
\author{Vitor V.~Lopes}
\affiliation{Energy Systems Modeling and Optimization Unit,
        National Laboratory for Energy and Geology (LNEG),
        Estrada do Pa\c{c}o do Lumiar 22, 1649-038 Lisbon, Portugal}
\author{Pedro G.~Lind}
\affiliation{Center for Theoretical and Computational Physics,
        University of Lisbon, Av.~Prof.~Gama Pinto 2,
        1649-003 Lisbon, Portugal}
\affiliation{ForWind - Center for Wind Energy Research, Institute of Physics,
Carl-von-Ossietzky University of Oldenburg, DE-26111 Oldenburg, Germany}

\date{\today}

\begin{abstract}
Using a method for stochastic data analysis, borrowed from statistical physics, we analyze synthetic data from a Markov chain model that reproduces measurements of wind speed and power production in a wind park in Portugal. 
We first show that  our analysis retrieves indeed the power performance curve, which yields the relationship between wind speed and power production and we discuss how this  procedure can be extended for extracting unknown functional relationships between pairs of physical variables in general.
Second, we show how specific features, such as the rated speed of the wind turbine or the descriptive wind speed statistics, can be related with the equations describing the evolution of power production and wind speed at single wind turbines.
\end{abstract}



\maketitle

\section{Introduction}
\label{sec:int}
    The use of efficient and clean renewable energy sources is one of the major conditions required to achieve the important aim of sustainable development in modern societies\cite{sustainable}. Wind energy is one of such sources and wind turbines are being subject to intensive studies for improving their efficiency\cite{windenergy,rauh04}. Although the basic laws of atmospheric wind motion have been known for a long time, important problems such as turbulence, layering, and the statistics of extreme events remain poorly understood. A better understanding of these phenomena can help to construct energy conversion schemes that are both more efficient and robust. Here,  robustness must be considered under two aspects: first, the occurrence of sudden changes in wind speed and direction can interrupt the process of energy conversion, meaning unreliability and a sudden slump in the electrical energy generated, which is seen as one of the major obstacles for the replacement of fossil and nuclear plants by
wind energy sources. Second, these sudden changes introduce massive mechanical stresses which can lead to excessive wear or, ultimately, to the destruction of wind generators.

    \begin{figure}[htb]
        \centering
        \includegraphics[width=0.48\textwidth]{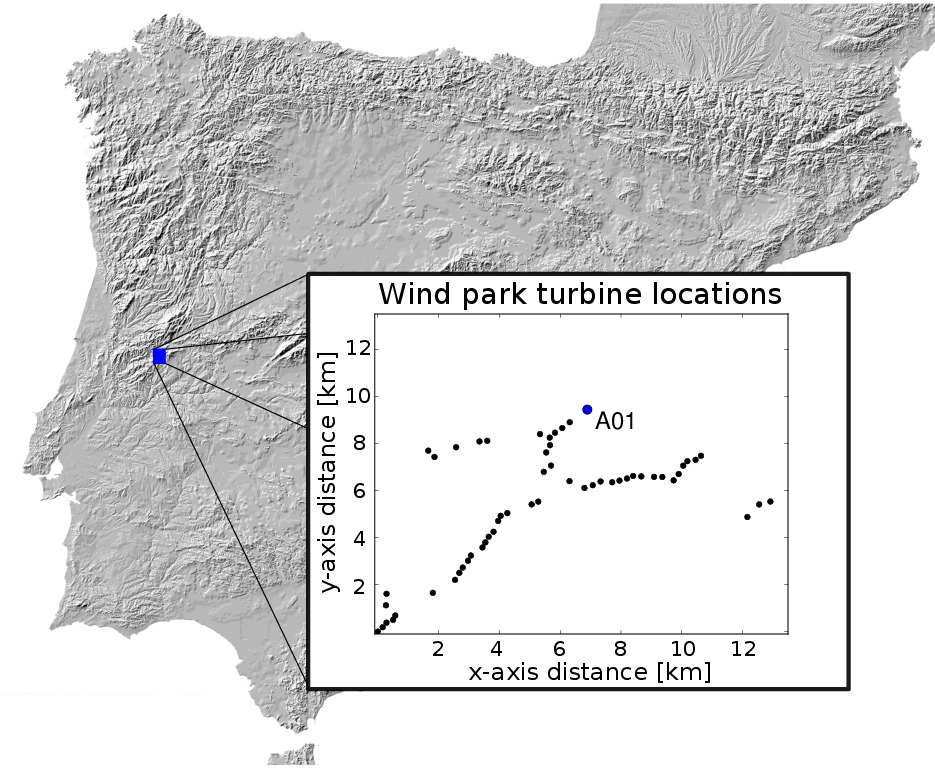}
        \caption{\protect Illustration of Iberia Peninsula indicating the position of the Portuguese wind park. In the inset one sees  the geographic location of each of the $57$ wind turbines (bullets). The blue marker (AO1) indicates the wind turbine analyzed here.}
        \label{fig1}
    \end{figure}

    Wind flow is in general turbulent\cite{milan2013} and non-homogeneous\cite{gottschall2007} with a non-negligible stochastic contribution. Therefore, in order to be able to construct more accurate models for its physical properties, one needs either accurate measurements of the wind speed on the length scales of wind turbines, or suitable models that can statistically reproduce these measured data. Since the wind turbines are driven by turbulent wind fields, the stochasticity of the wind fields transfers to stochastic dynamics of the wind turbine as a whole, of the loads on its structures and, last but not least, of the power output. Recently, a Markov chain model was used to reproduce wind measured data\cite{Lopes2012}, based on the transition matrix and time propagators for the wind speed and direction together with the power production. Differently from previous first-order approaches\cite{Sahin2001}, information from two and three step transition probabilities are considered.

    \begin{figure}[h!]
        \centering
        \includegraphics[width=0.46\textwidth]{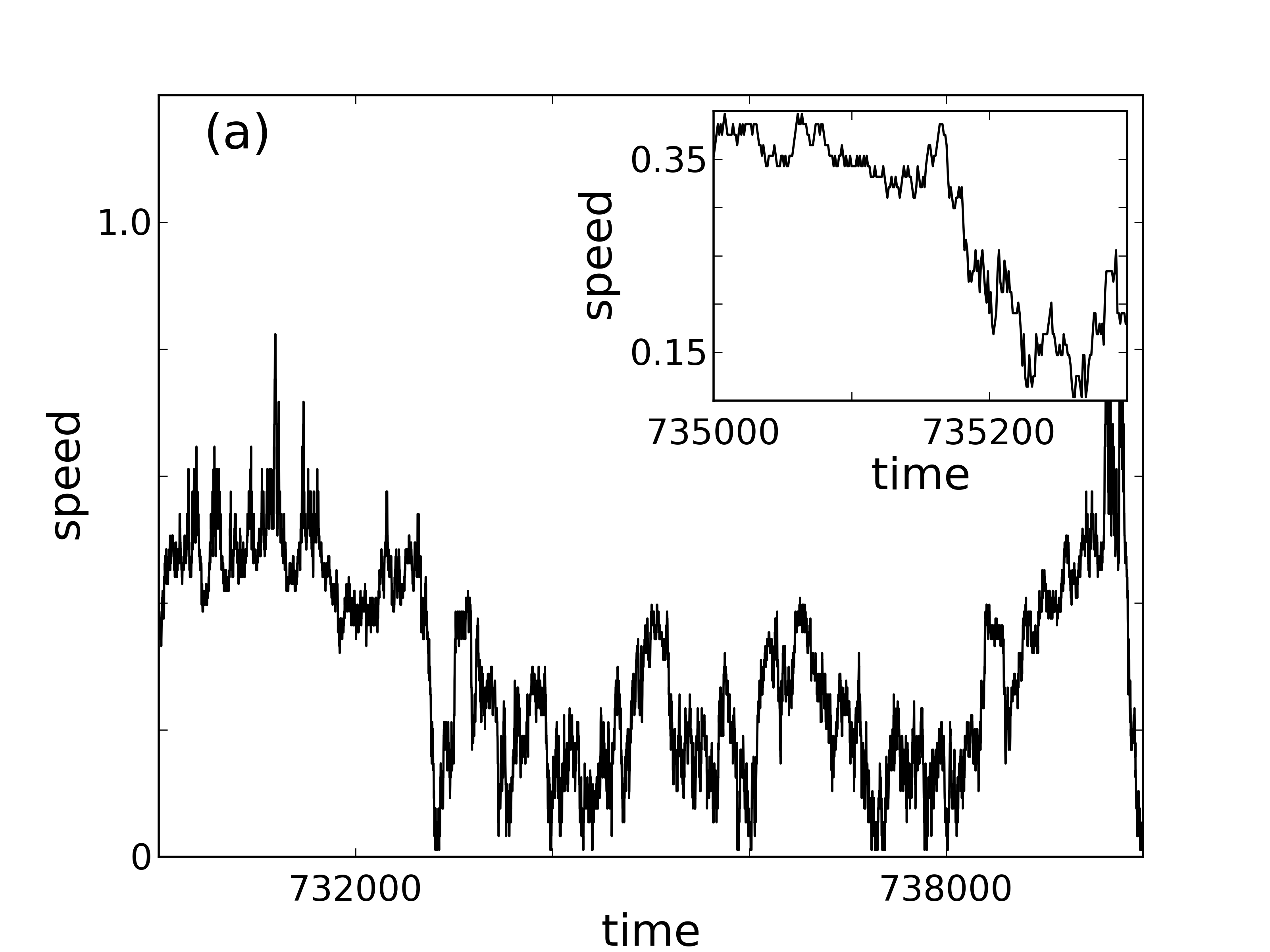}
        \includegraphics[width=0.46\textwidth]{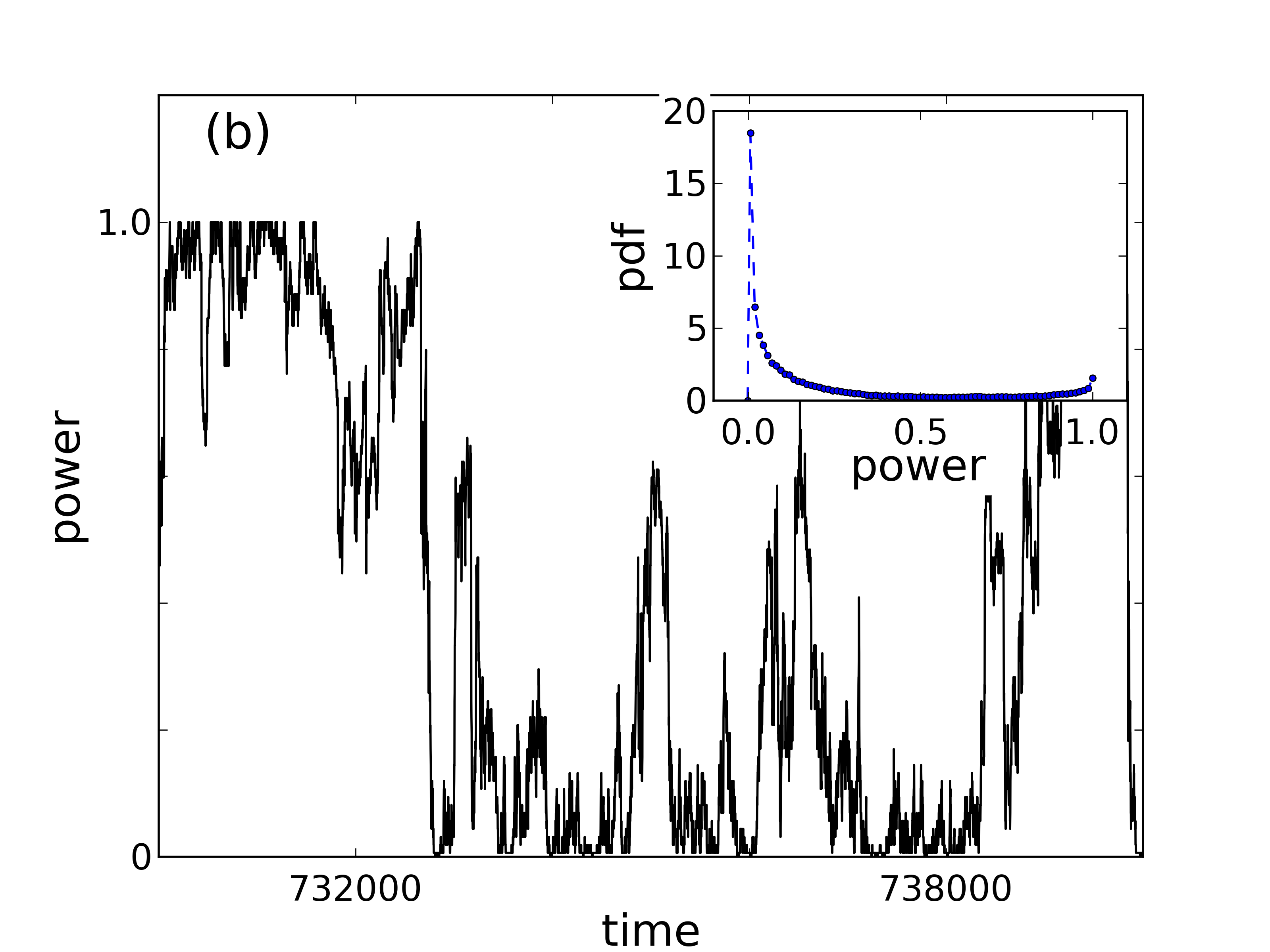}
        \includegraphics[width=0.46\textwidth]{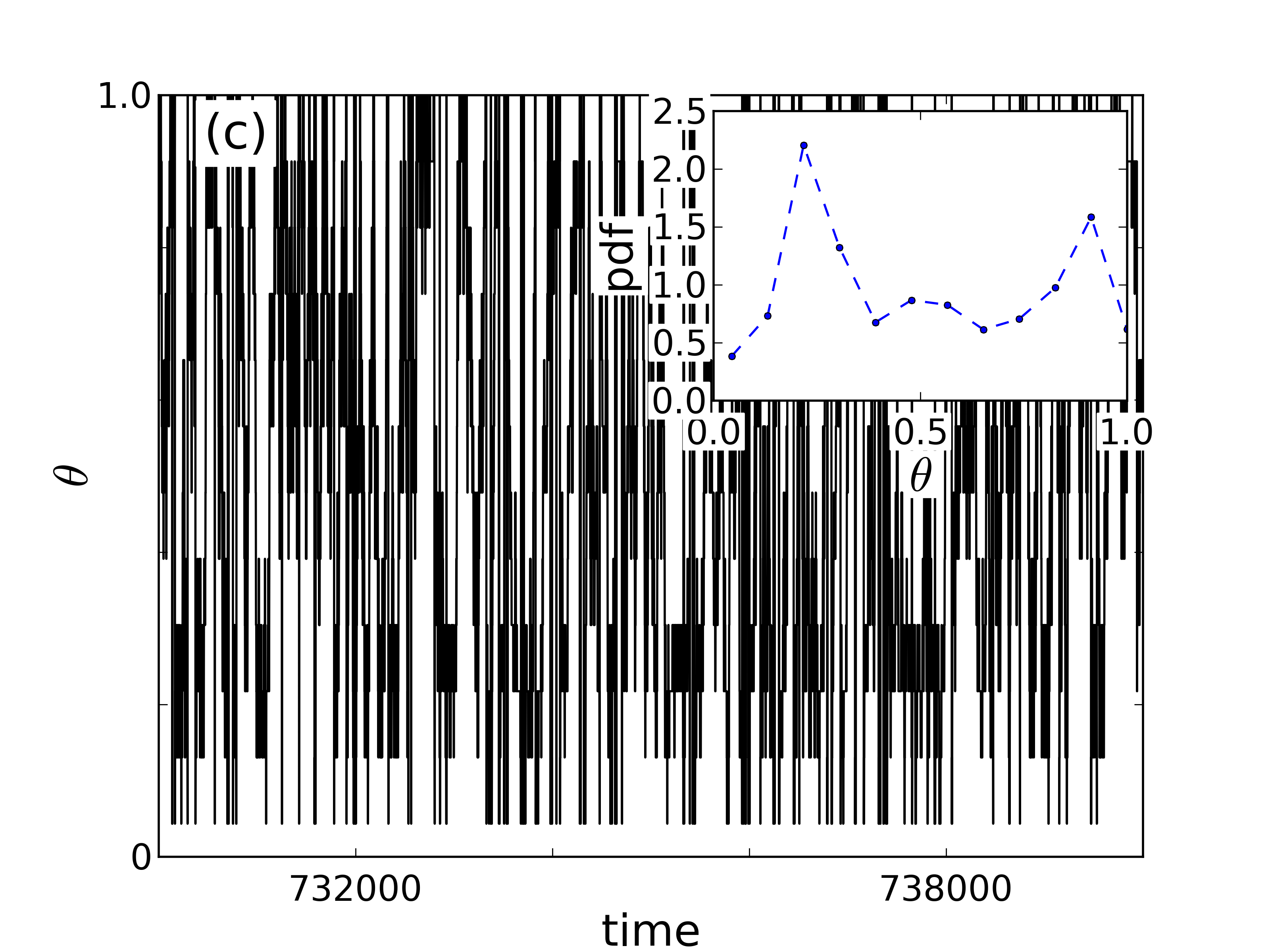}
        \caption{\protect Time series for 
     {\bf (a)} the magnitude of the wind speed $v$. Inset shows the  time series for a shorter time period. The PDF of the wind speed is shown in the inset of Fig.~\ref{fig5}.
     {\bf (b)} the power production $P$ of the wind turbine. Inset shows the PDF of the time series.
     {\bf (c)} the corresponding wind direction, $\theta$. Inset shows the PDF of the time series.
All data series were generated with the Markov chain model \cite{Lopes2012} described in Sec.~\ref{sec:data}. All properties are normalized to the observed intervals 
     $[0,v_{max}]$, $[0,P_{max}]$ and $[0,\theta_{max}]$ respectively.
     In this and all following figures, time is in multiples of 20min, whereas velocity $v$, power $P$ and direction $\theta$ are normalized to unity.}
        \label{fig2}
    \end{figure}

    \begin{figure}[htb]
        \centering
        \includegraphics[width=0.5\textwidth]{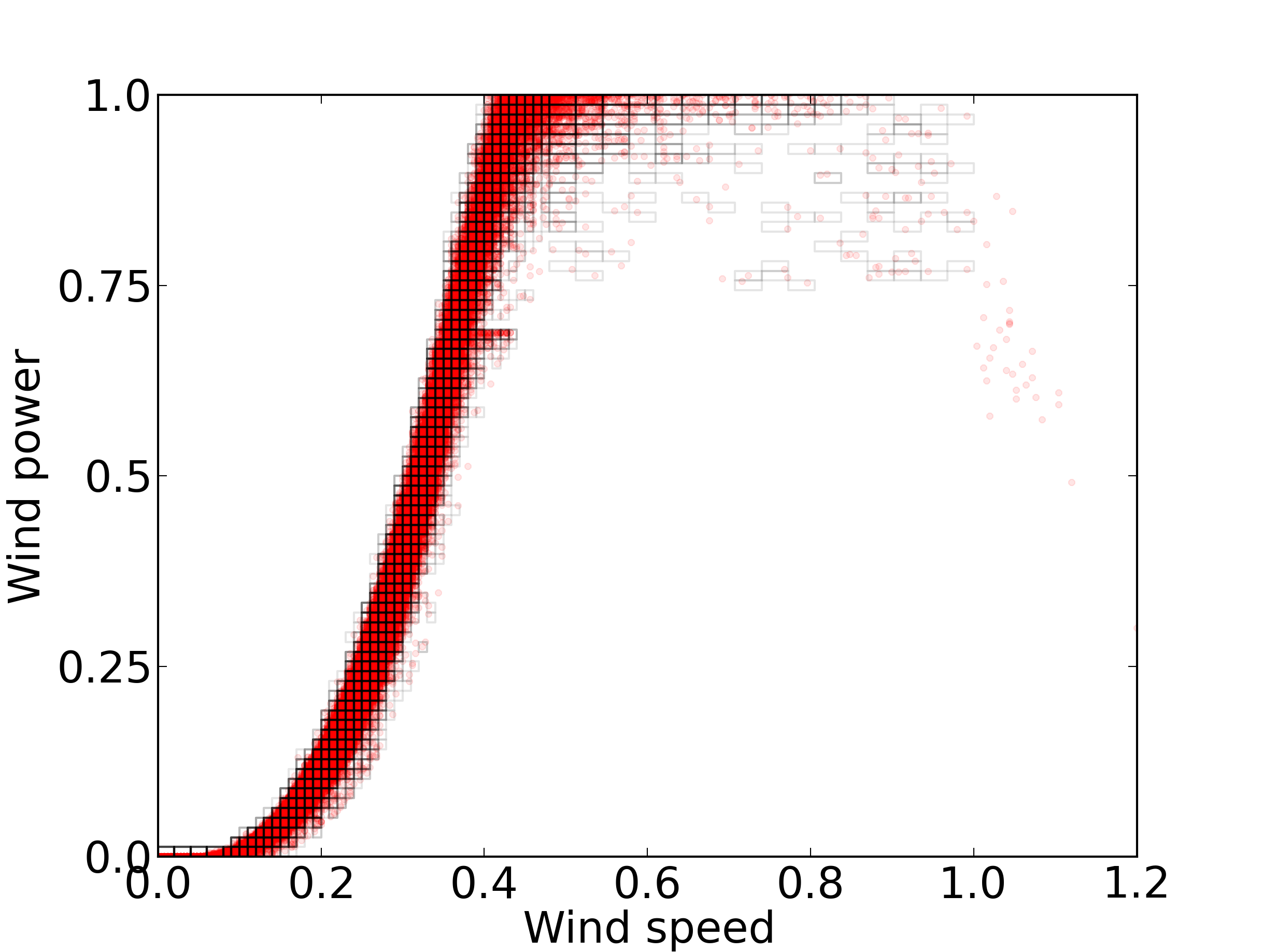}
        \caption{\protect (Color online) Performance curve for one wind turbine in Pinhal Interior, Portugal. Circles show all the historical data points used in the Markov chain modeling \cite{Lopes2012} and the boxes display the state discretization.} 
        \label{fig3}
    \end{figure}

    In this paper, we aim at understanding the stochastic aspects of power production coupled to the wind velocity field. To that end, we use a methodology recently introduced by some of us\cite{vitor}, for uncovering optimal stochastic variables weakly\cite{russo12} and strongly coupled\cite{vitor}, and adapt it with two purposes. First, to properly derive the functional relation of pairs of variables whose values are extracted from the Markov chain model for wind turbines. Second, to uncover specific features of the wind turbine and characterize the different working regions observed in the power-speed plane. After this, we test our approach to uncover the functional dependence of the well-known performance curve, which describes the functional dependence of the power production and the wind speed. Whereas previous reports have pointed out the benefits of deriving the power curve from the drift field\cite{anahua2008}, we additionally take the diffusion field into account and find that this procedure 
creates 
additional insight.
 
We use the data sets generated by the Markov chain model described by Lopes et al.~\cite{Lopes2012}. Using such synthetic data sets that properly reproduce the statistical features of empirical data sets, allows us to use data sets as large as needed for our analysis. Moreover, the Markov chain model  serves as a  filter to remove periodicities present in the data. This is  an essential step for our modeling, as  it would not be possible to correctly estimate the Kramers-Moyal coefficients from the raw data. 

    We start in Sec.~\ref{sec:data} by describing the empirical data used to define the Markov chain model as well as the data generated with it. In Sec.~\ref{sec:analysis} we describe our stochastic method for analyzing the data and in Sec.~\ref{sec:performance}  we apply it to analyze the performance curve of a wind turbine. 
In Sec.~\ref{sec:integration} we show that the same method when applied 
separately to both wind speed and power production allows to derive
the performance curve. 
Further, the same analysis also provides insight concerning specific
features of the  turbine system studied.
Section \ref{sec:conclusions} concludes this paper.

\section{Properties and generation of the data sets}
\label{sec:data}

    The data analyzed in this manuscript was simulated from a set of measurements from a wind turbine in the region of Pinhal Interior, Portugal. The measured properties are the power production $P$ of the wind turbine, the wind speed $v$ and the wind direction $\theta$ ($\theta = 0$ corresponds to north).
The wind turbine was selected  out of a total of 57 wind turbines in an eolic park. Figure \ref{fig1} shows an overview of the eolic park. The time increment between two successive measures is $\Delta t=10$ minutes and the time period covered starts in January 1st 2009 and ends in December 31st 2010, yielding approx.~$10^5$ data points. It has to be remarked that these measurements are acquired directly from the top of wind turbine (nacelle) and might not be optimal for the reconstruction of the underlying physical processes for two reasons: first, the wind speed measurement is acquired at a  point located downstream of the turbine blades and can neither account for the spatial extension and inhomogeneity of the wind field nor for its complex aerodynamical interaction with the turbine blades\cite{matthias2012}. Secondly, the 10 minute sampling period of the historical data set does not allow to resolve the time scales of either the turbulent interaction between wind and turbine, nor the quick 
action of the controller system response. Finally, missing data records, a low number of data points \cite{david_mle} or large sampling intervals \cite{lade09b}, and periodicities due to the daily and seasonal variations in wind flow often hinders a direct stochastic analysis of these data sets. %
Specifically,  it is well known  that the 
estimation of the drift and diffusion coefficients which we use cannot 
be applied to periodic time series. If applicable, a  filtering or detrending procedure has to be applied to the data set \cite{russo12, fpeq, gradisek_eigenvectors}. 
%
%
However, as the  quasi-daily variations  in the wind speed do not occur every day, the usefulness of a filtering or detrending procedure has to be doubted.

Thus, the challenge is to devise new methods that can make most use of the information present on these data sets---given that most of the data acquisition systems on existent wind farms are limited---with the aim of  understanding  the dynamic processes of the wind power generation, which hopefully  can lead to  economic benefits from scheduling and maintaining a level of constant production.

   To overcome some of these problems, we employ a reconstruction of the original wind data set through a Markov chain model, which has recently been  established \cite{Lopes2012} using a joint discretization of the wind speed, power and direction variables for the state definition. In our specific case within the range of each variable $P$, $v$ and $\theta$ we select $80$, $60$ and $12$ states. States without a realization in the time series are deleted, all states that contain at least one observation are kept. Details about the estimation of the transition probabilities can be found in \cite{Lopes2012,Hocaoglu2008} 
and Appendix~\ref{append:Markovmodel}. %
The use of the MC model has the additional benefit of removing periodicities from the data.

Based on the Markov chain transition matrix $\mathbf{P}$ with $\mathbf{P}(i, j) = p_{i,j}$ being the probability of transition from state $s_i$ to state $s_j$, the synthetic data sets were generated using the following Monte Carlo approach. We find the cumulative probability transition matrix $\mathbf{P}_{cum}$ with $\mathbf{P}_{cum}(i, j) = \sum_{k = 1}^{j}p_{i,k}$ and select randomly an initial state $s_{i}$. A random number $\epsilon$ between zero and one is then uniformly selected and a new state $s_{new}$ is chosen such that $ P_{cum}(i, {new}) \geq \epsilon.$ For details see Ref.~\cite{Sahin2001}. Figure \ref{fig2} shows the generated time series.
As shown in the previous work Ref.~\cite{Lopes2012}, the proposed Markov chain model reproduces the dominant statistical features of all three properties, namely power production,
wind speed and wind direction, although no periodicities are present in the reproduced time series.

The resulting synthetic data series for power production and wind speed also retain the persistence statistics, namely the average duration of power production and wind speed on a certain level, respectively. Power production and wind speed are presented as fractions of the maximum observed power $P_{max}$ and wind speed $v_{max}$ respectively, assuming therefore values between zero and one.

\begin{figure*}[htb]
        \centering
        \includegraphics[width=0.45\textwidth]{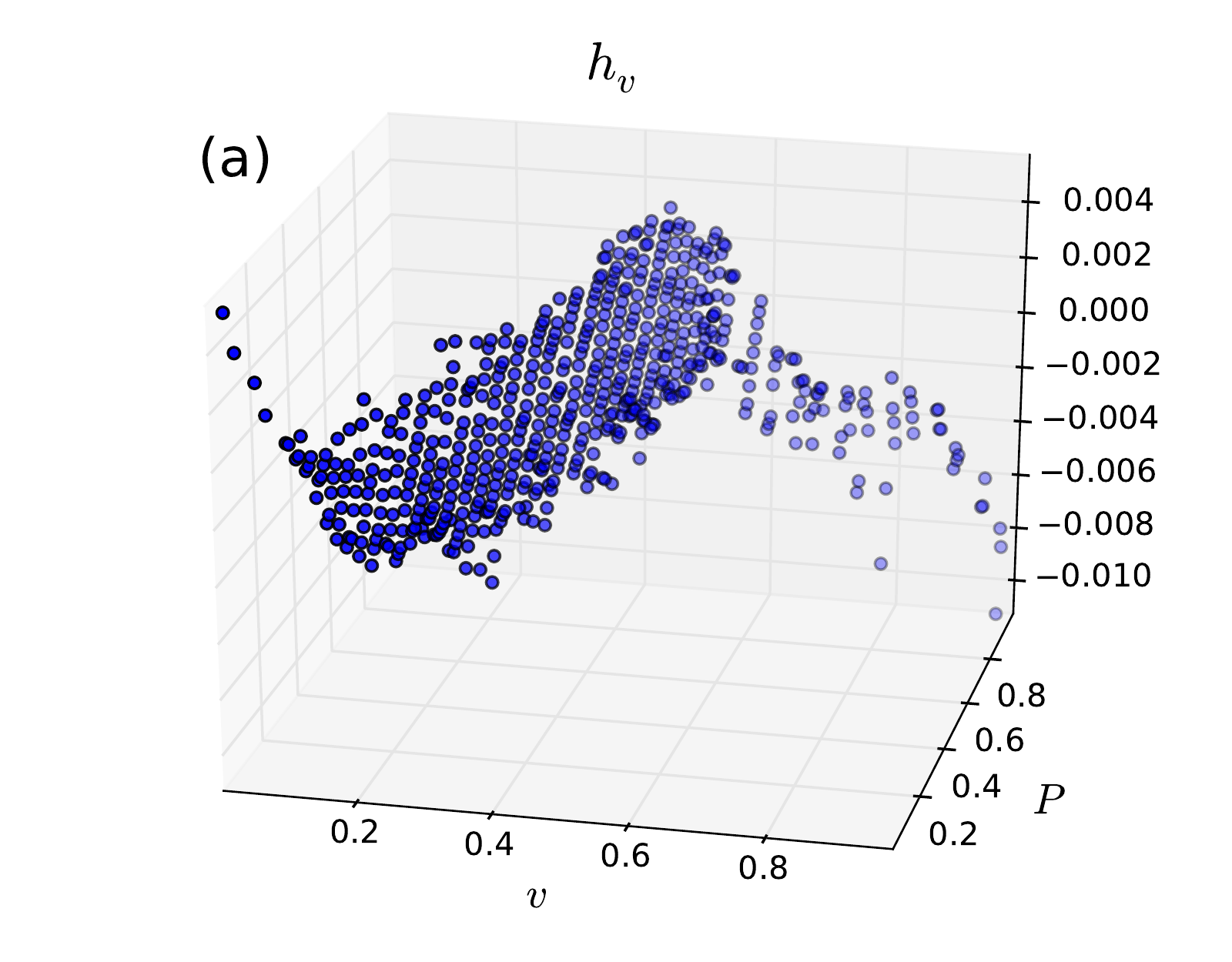}%
        \includegraphics[width=0.45\textwidth]{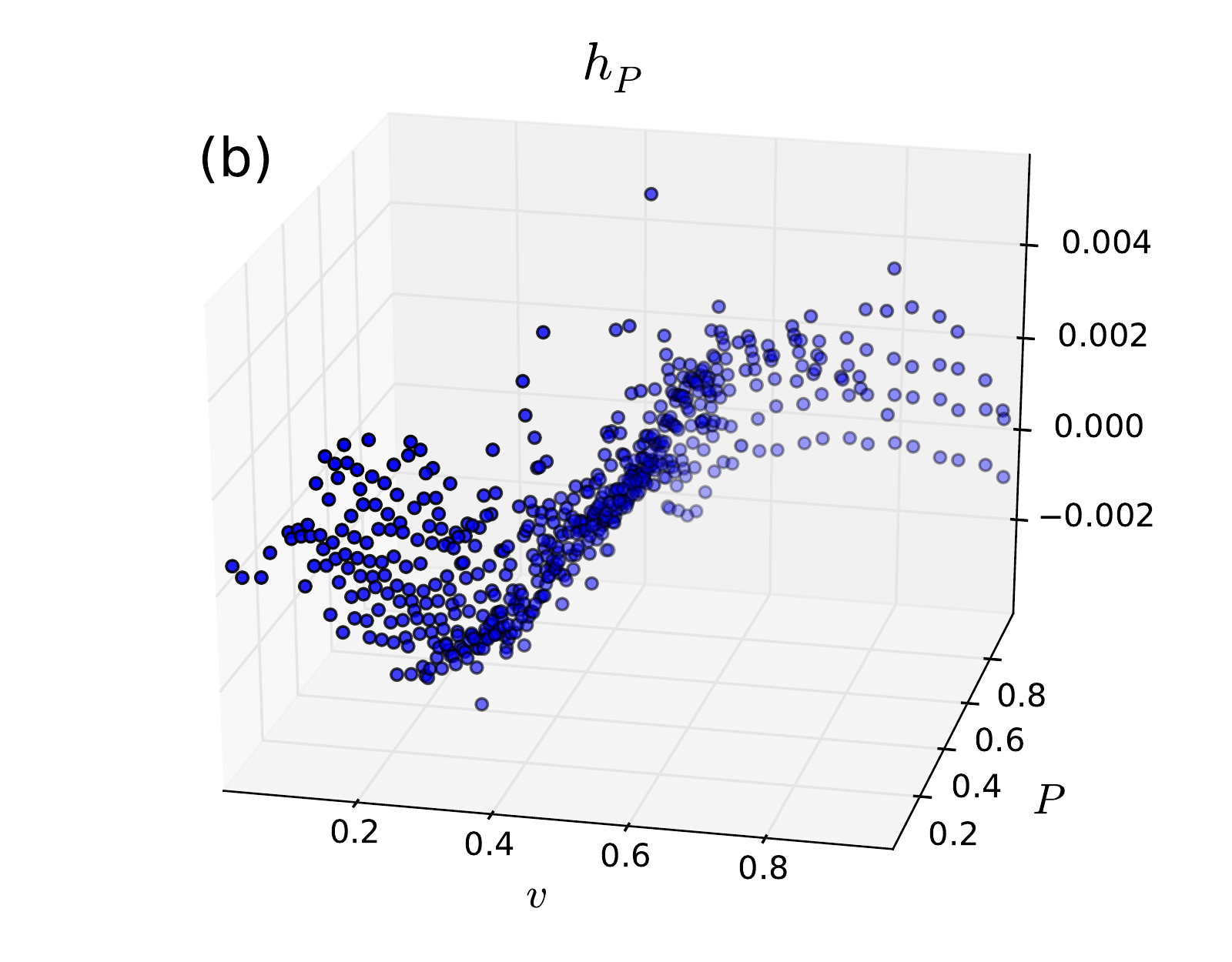}
        \includegraphics[width=0.45\textwidth]{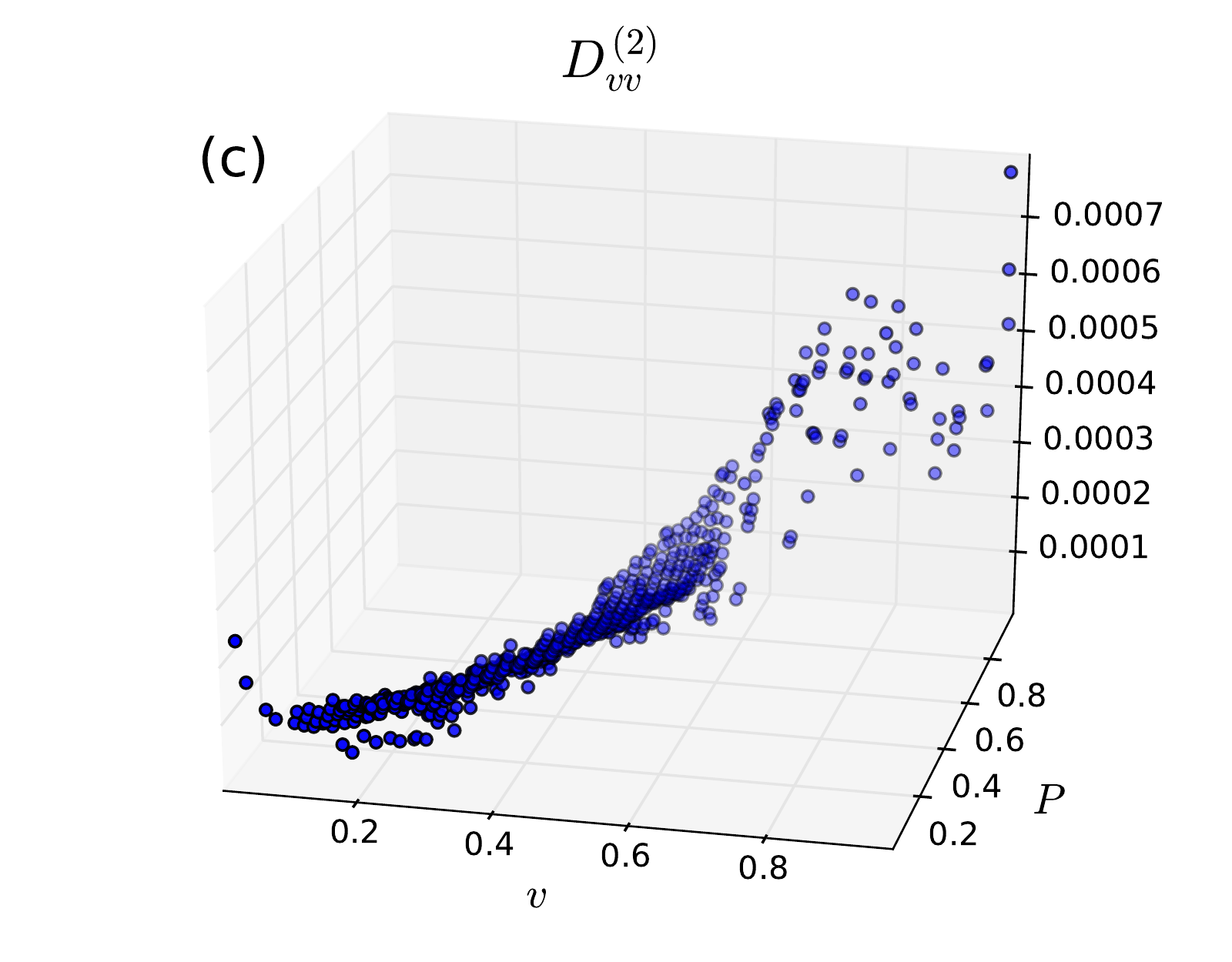}%
        \includegraphics[width=0.45\textwidth]{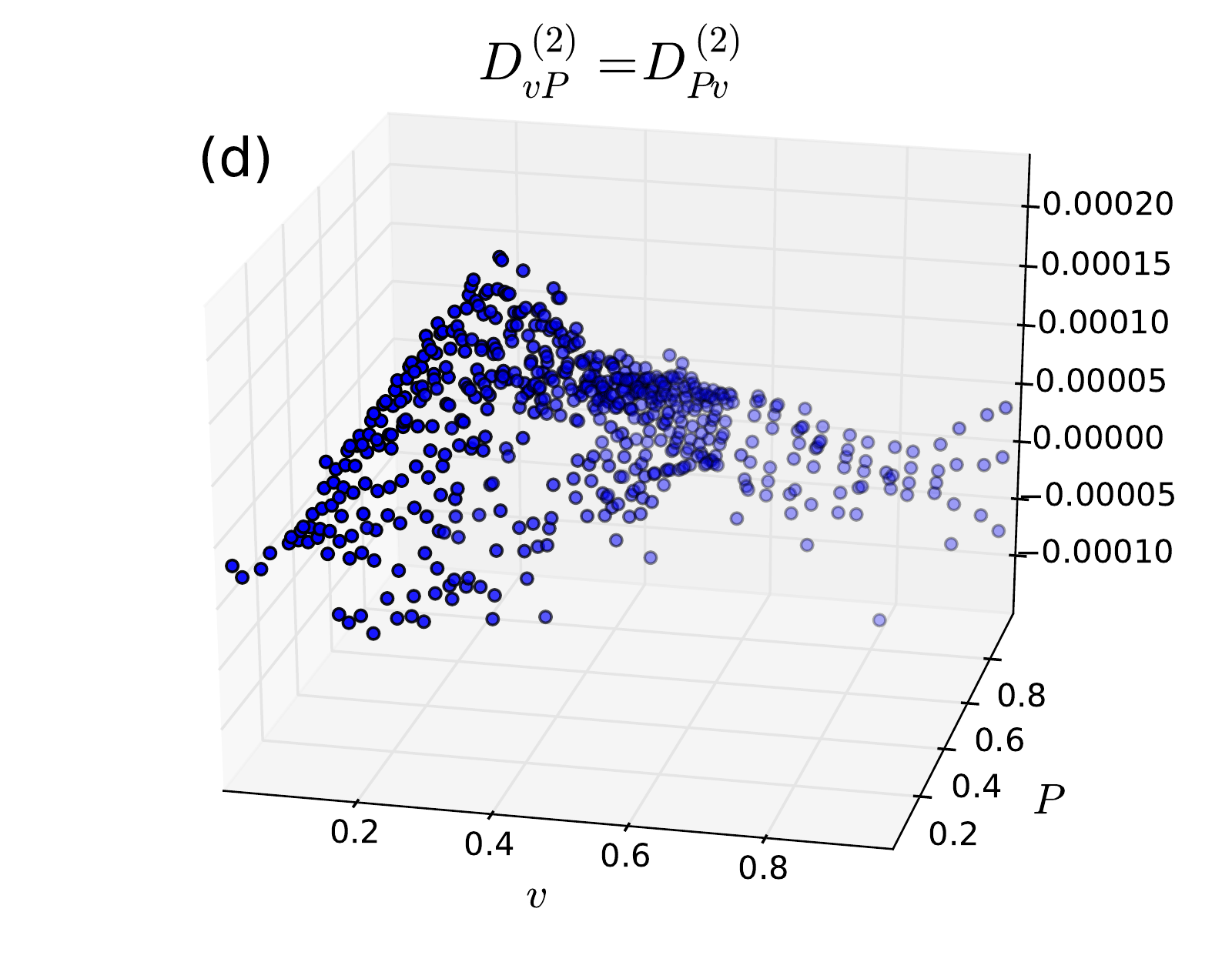}
        \includegraphics[width=0.45\textwidth]{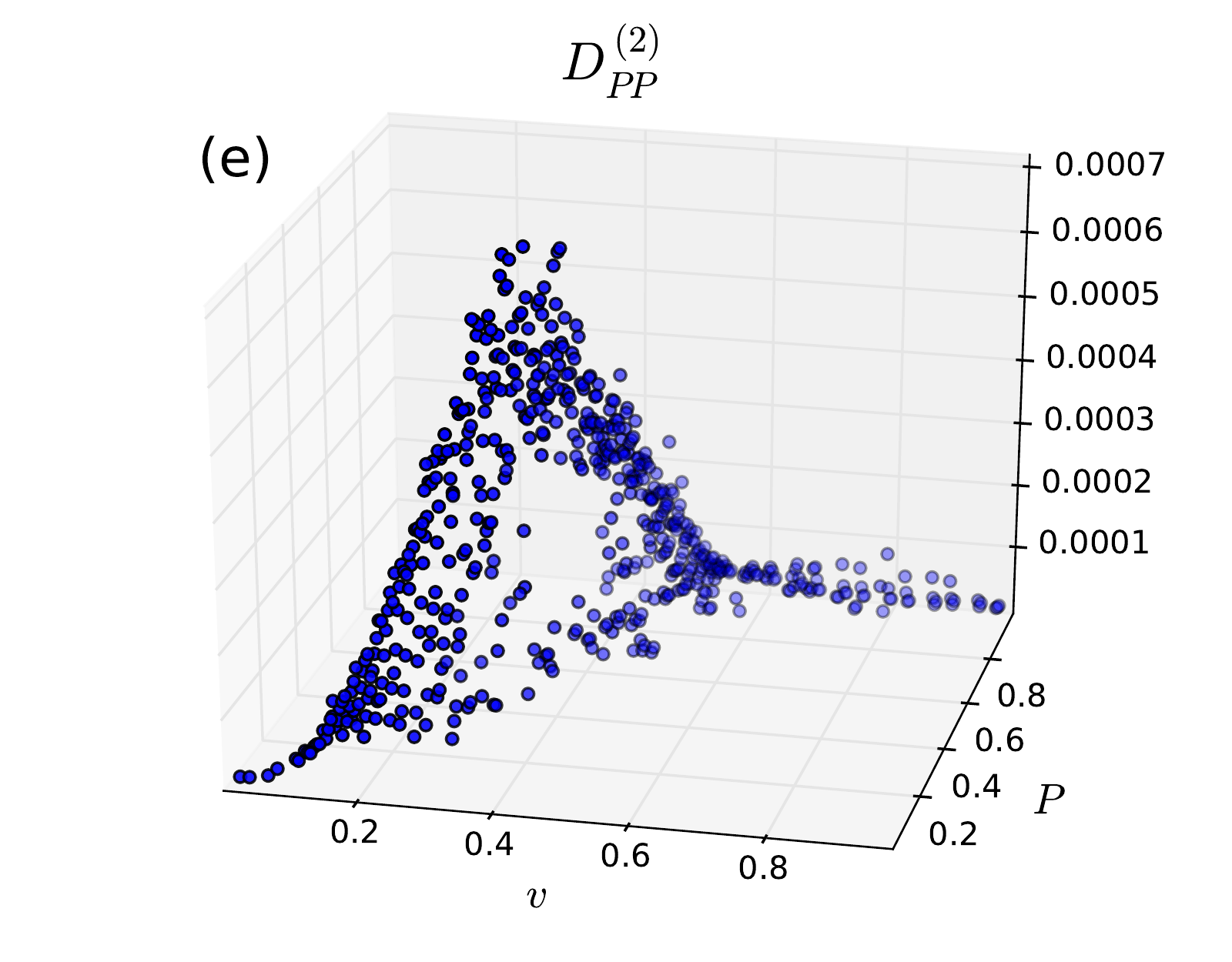}%
        \includegraphics[width=0.45\textwidth]{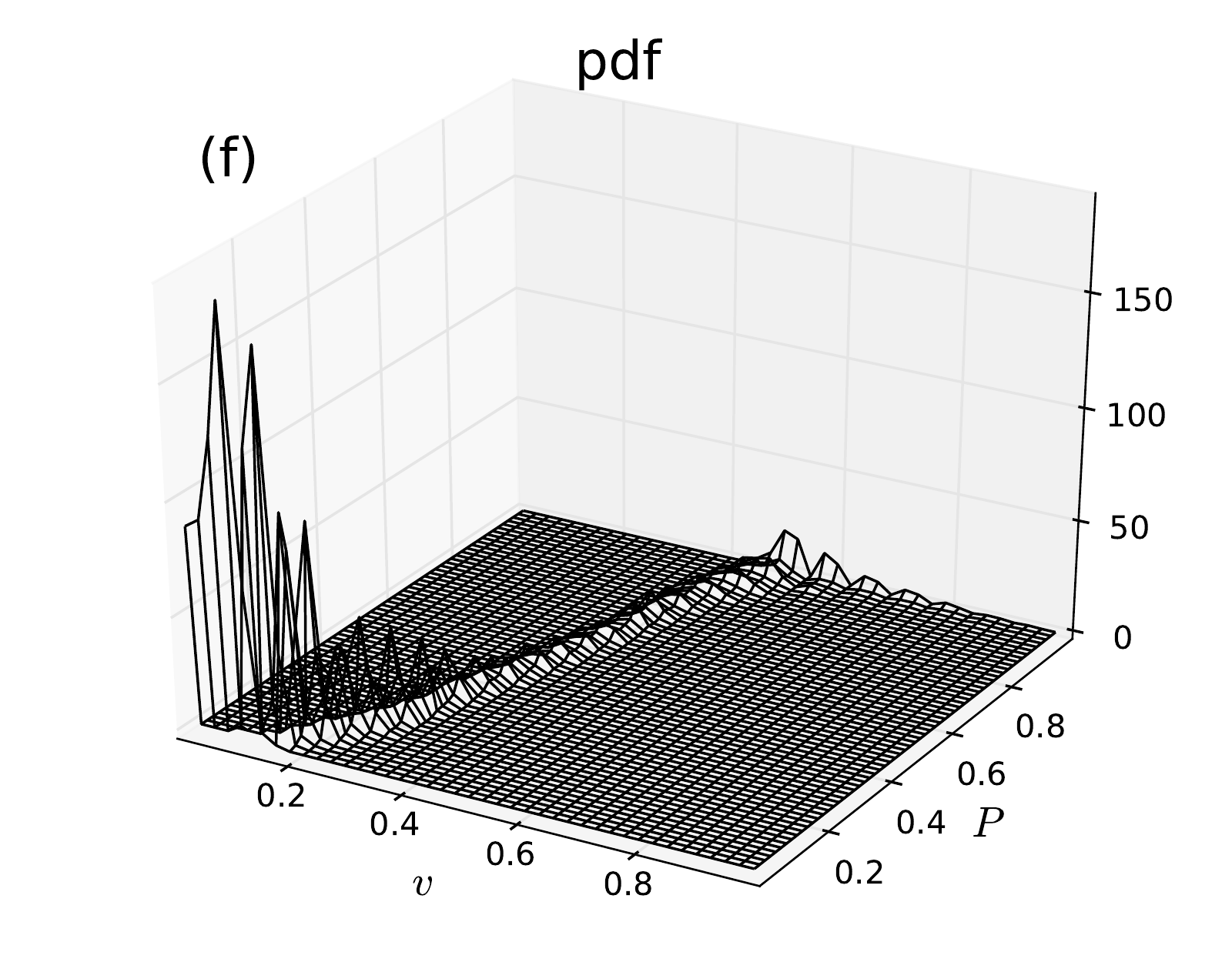}
        \caption{\protect (Color online) The drift and diffusion coefficients defining the co-evolution of $P$ and $v$: {\bf (a)} $h_{v}$, {\bf (b)} $h_{P}$, {\bf (c)} $D_{vv}^{(2)}$, {\bf (d)} $D_{vP}^{(2)}=D_{Pv}^{(2)}$ and {\bf (e)} $D_{PP}^{(2)}$. The PDF of both variables is shown in {\bf (f)}.}
        \label{fig4}
    \end{figure*}

The wind direction in the original time series follows a bimodal pattern, which is to some extent also periodic:  during the day there are weaker breezes in a particular direction than during the night when wind streams in a different direction; therefore the prevailing wind speed values occurring during the ``day'' (2pm -- 2am) have approximately $\theta_1\sim 80^{\circ}$, a value different from the one found for wind speeds measured during the ``night'' (2am--2pm), $\theta_2\sim 320^{\circ}$. The Markov chain model is capable of preserving this bimodality. Although we will not consider the wind direction in our study,  one should notice that the wind direction bimodality is reflected in the distribution of the wind speed (see inset of Fig.~\ref{fig5} below). Another bimodality preserved from the original data set is the one in the joint PDF of velocity and power, which is visible in Fig.~\ref{fig4}f.

Figure \ref{fig3} shows the discretization of the data set where the combined states for power and speed are indicated with boxes and circles represent the historical data set points projected into the speed-power plane.
The synthetic data sets were generated with $2\times 10^6$ data points and show stationary behavior, i.e.~have constant moving averages (not shown).

This approach has several advantages over the direct analysis of historical data sets. First, high-quality data series of arbitrary length can be generated, which increases the accuracy of the Markov analysis. Secondly, the generated data are by construction Markovian, with the reconstruction through the Markov chain acting as a filter that removes both noise correlations and periodicities. Finally, non-Gaussian transition probabilities between the states are preserved, which enables to study them through higher Kramers-Moyal (KM) coefficients. Error analysis for the derivation of the transition matrix is described
in Append.~\ref{append:errors}.

    \begin{figure*}[htb]
        \centering
        \includegraphics[width=1.00\textwidth]{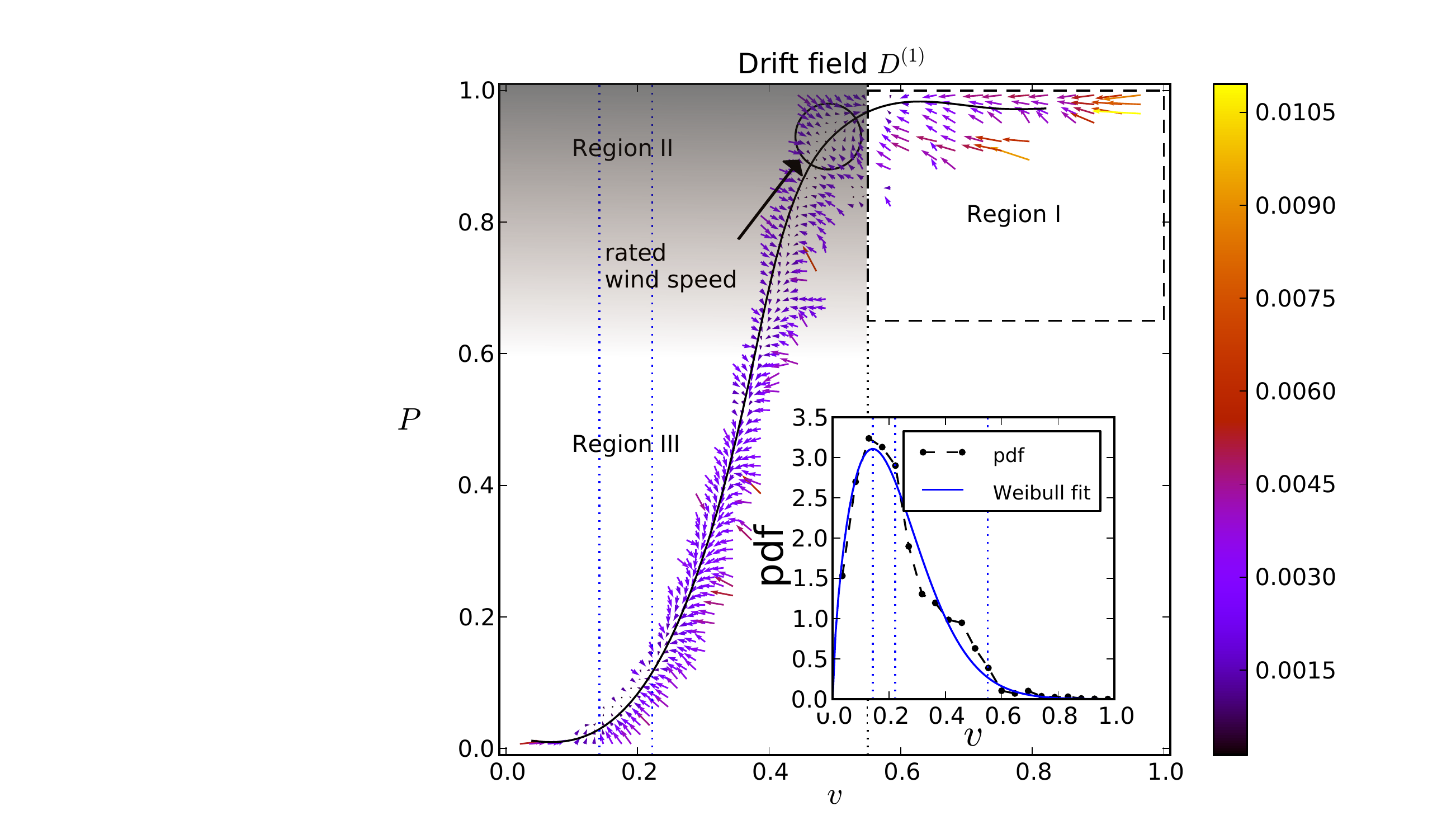}%
        \caption{\protect (Color online) The drift vector $\f{D}^{(1)}(P,v)=(h_P(P,v),h_v(P,v))$ (see Eq.~(\ref{power-velocity})) in each $(P,v)$ box used to generate the data (see Fig.~\ref{fig3}). Three regions can be identified: Region I having slow dynamics and regions II and III with fast dynamics (see text). Interestingly, the fixed point spot in Region II coincides with the rated speed of the wind turbine (see text). The black curve indicates the performance curve and crosses the bins for which $D^{(1)}(v,P)$ vanishes. In the inset, the marginal probability density function of the wind speed $v$ is well fitted by a Weibull distribution with  
scale parameter $\lambda \approx 0.25$ and shape parameter $k \approx 1.66$ and a mode $v_{mode} \approx 0.14 v_{max}$\cite{weibull}. 
Velocities above $v_{th}=0.55v_{max}$ are rarely observed, which explain the observed drifts in Region I (see text). The gray shading indicates the gradual transition from Region II to Region III. Dotted vertical lines mark the positions of $v_{mode}$, $\langle v\rangle$ and  $v_{th}$, respectively.}
        \label{fig5}
    \end{figure*}

\section{Stochastic analysis of wind turbines}
\label{sec:analysis}

    The co-evolution of two or more stochastic variables, such as wind speed and power production, can be described through a system of coupled stochastic equations, each one defined by a deterministic contribution (drift) and stochastic fluctuations from possible stochastic sources. In this section we present the general framework to analyze our data and in the next section we apply it to the power production and wind speed variables.

    For the general case of $K$ stochastic variables, $X_1,\dots,X_K$ the vector  ${\bf X}(t)=(X_1(t),...,X_K(t))$ defines the state of the system under study at each time instant $t$. The evolution of the state vector yields a stochastic trajectory in phase space and is given by the so-called It\^o-Langevin equation\cite{vitor,fpeq,gard}:

    \begin{equation}
        \frac{d \mathbf{X}}{d t}=
                    \mathbf{h}(\mathbf{X})
                    + \mathbf{g}(\mathbf{X})
                    \mathbf{\Gamma}(t) ,
        \label{Lang2DVect}
    \end{equation}
    where $\mathbf{\Gamma(t)}=(\Gamma_1(t),\dots,\Gamma_K(t))$ is a set of $K$ independent stochastic forces with Gaussian distribution fulfilling the following conditions: $\langle \Gamma_i(t)\rangle = 0$ and $\langle \Gamma_i(t)\Gamma_j(t')\rangle = 2\delta_{ij}\delta(t-t')$. Function $\mathbf{h}=\{ h_i\}$ in Eq.~(\ref{Lang2DVect}) is the deterministic contribution, describing the physical forces which drive the system, while $\mathbf{g}=\{ g_{ij} \}$ describes the amplitude of the stochastic sources of fluctuations $\mathbf{\Gamma}$\cite{physrepreview}.

    The evolution of the stochastic variables in time yields a joint probability density function (PDF), $f(\mathbf{X})$, 
    that evolves according to the so-called Fokker-Planck equation
    \begin{eqnarray}
        \frac{\partial f(\mathbf{X},t)}{\partial t} &=&
        -\sum_{i=1}^N\frac{\partial}{\partial x_i}
            \left [
            D_i^{(1)}(\mathbf{X})f(\mathbf{X},t)
            \right ] \cr
        & &+\sum_{i=1}^N\sum_{j=1}^N
            \frac{\partial ^2}{\partial x_i\partial x_j}
            \left [
            D_{ij}^{(2)}(\mathbf{X})f(\mathbf{X},t)
            \right ] \quad ,
        \label{FPE}
    \end{eqnarray}
    where the functions $D_i^{(1)}$ and $D_{ij}^{(2)}$ are related to the functions $h_i$ and $g_{ij}$ above, namely
    \begin{subequations}
        \begin{eqnarray}
            D_i^{(1)}(\mathbf{X}) &=& h_i (\mathbf{X}) \label{D1}\\
            D^{(2)}_{ij}(\mathbf{X}) &=& \sum^N_{k=1}g_{ik}({\bf X})g_{jk}({\bf X})\label{D2}
        \end{eqnarray}
    \label{KMs}
    \end{subequations}
    and are usually called drift and diffusion functions, respectively.

    Drift and diffusion functions can be directly derived from observed or generated data \cite{physrepreview,lind10}, and this fact is the basis of our framework. Indeed, the drift and diffusion functions of the underlying process are defined through conditional moments, namely \cite{fpeq}:

    \begin{equation}
        \mathbf{D}^{(k)}(\mathbf{X})=\lim_{\Delta t\rightarrow0}\frac{1}{\Delta t}
                                \frac{\mathbf{M}^{(k)}(\mathbf{X},\Delta t)}{k!}
        \label{DefCoefKM}\quad,
    \end{equation}

    where $\mathbf{M}^{(k)}$ are the first and second conditional moments ($k=1,2$). These conditional moments can be directly derived from the measured data as \cite{physrepreview,lind10} $M_i^{(1)}(\mathbf{X},\Delta t) = \left\langle Y_i(t+\Delta t)-Y_i(t)  | {\mathbf{Y}(t)=\mathbf{X}} \right\rangle $ and $M_{ij}^{(2)}(\mathbf{X},\Delta t) = \left\langle (Y_i(t+\Delta t)-Y_i(t))(Y_j(t+\Delta t)-Y_j(t)) | {\mathbf{Y}(t)=\mathbf{X}}\right\rangle  $ where $\mathbf{Y}(t)=(Y_1(t),\dots,Y_N(t))$ is the $N$-dimensional vector of measured variables and $\langle \cdot | {\mathbf{Y}(t)=\mathbf{X}} \rangle$ symbolizes a conditional averaging over the entire measurement period, where only measurements with ${\mathbf{Y}(t)=\mathbf{X}}$ are taken into account. Important conditions to hold are (i) the underlying process is stationary and (ii) the Markovian property is fulfilled.

   Numerically $\mathbf{h}$ and $\mathbf{g}$ are determined on a $n_1\times...\times n_N$ mesh of points in phase space, as a function of the variables $X_i$, using the drift and diffusion functions. Locally, at each mesh point, one can always diagonalize the matrix $\mathbf{g} (\mathbf{X})$ and compute their  $K$ eigenvalues and $K$ eigenvectors. As shown previously\cite{vitor,gradisek_eigenvectors,vanMourik_eigenvectors}, this analysis provides information about the stochastic forces acting on the system. Namely, the eigenvalues indicate the amplitude of the stochastic force and the corresponding eigenvector indicates the direction toward which such force acts. In a previous work\cite{vitor} we argued that to each eigenvector of the diffusion matrix one can associate one independent source of stochastic forcing $\Gamma_i$ and thus the eigenvectors can be regarded as defining principal axes for stochastic dynamics. In particular, if one eigenvalue is very small compared to all the others, the corresponding 
stochastic force can be neglected. In the following sections we present a different implication of this principal stochastic component analysis, which emphasizes that the vanishing of one stochastic direction is in fact an indication of a strong functional dependence between the pair of variables being analyzed.

    \begin{figure*}[htb]
        \centering
        \includegraphics[width=.80\textwidth]{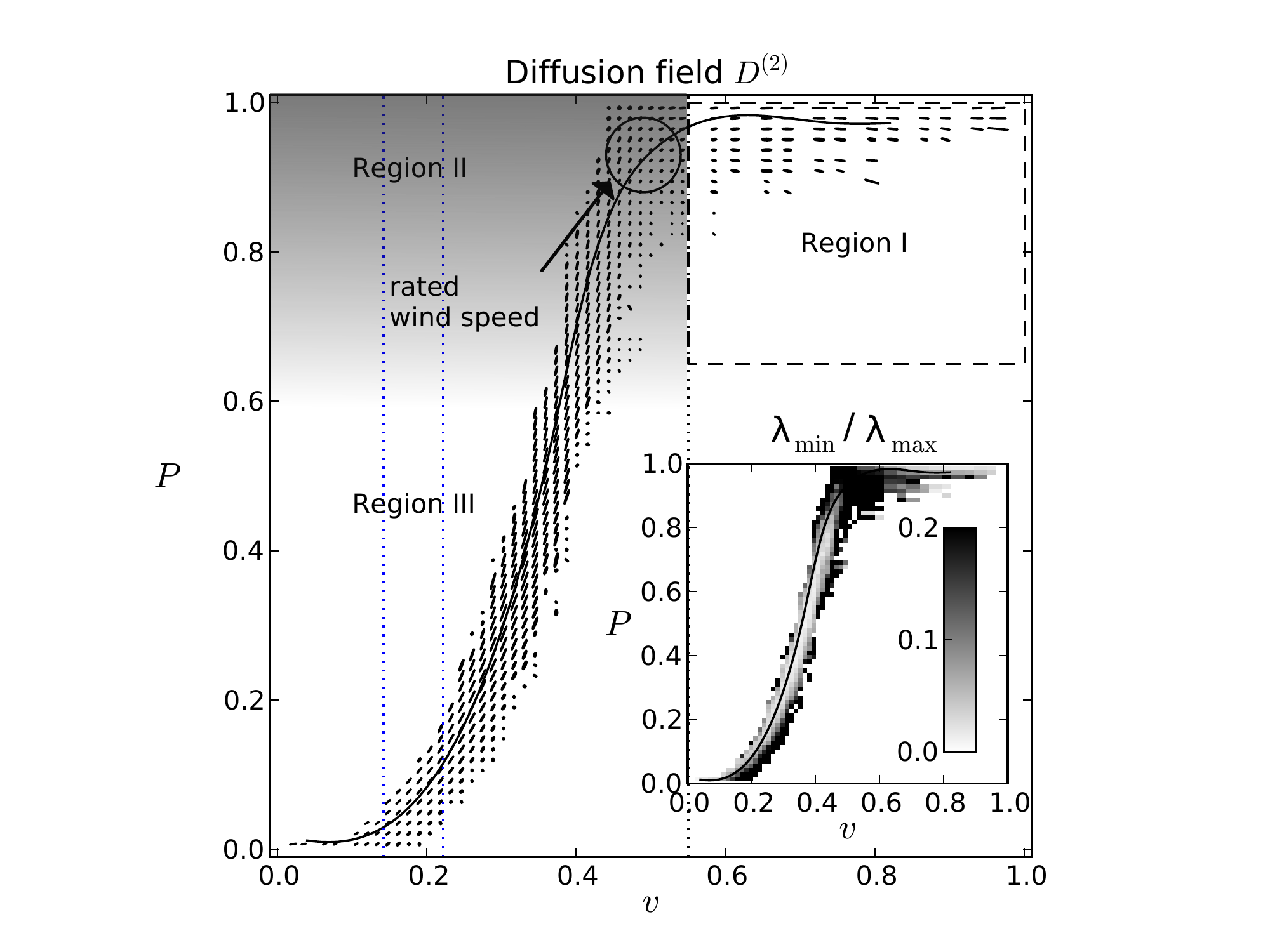}
        \caption{\protect 
(Color online) Diffusion ellipses in the power production and wind speed state space, plotted together with the distribution of data pairs along the boxes shown in Fig.~\ref{fig3}. At each box center, the corresponding diffusion ellipse is defined by the two orthogonal eigenvector of the diffusion matrix $\f{D}^{(2)}$ computed at that center. The principal axes defining the ellipse are aligned along the eigenvectors with a length proportional to the corresponding eigenvalue. Along the performance curve, which gives the functional  dependence between both variables, $P$ and $v$, the diffusion ellipses degenerate to a line segment tangential to the curve at each box center. This feature enables one to use the diffusion matrix of any set of variables for deriving their functional relationships (see text).  The gray shading indicates the gradual transition from Region II to Region III. In the inset one sees the ratio of  both eigenvalues $\lambda_{min}/\lambda_{max}$, using a gray scale ($0.2$ for black, $0$ for white). Dotted 
vertical lines mark the positions of $v_{mode}$, $\langle v\rangle$ and  $v_{th}$, respectively.}
        \label{fig6}
    \end{figure*}

\section{Wind turbine drift and diffusion map analysis}
\label{sec:performance}

In this section we focus solely on two variables, which are power production $P$ and wind speed $v$. Since both series are stationary and Markovian, we assume them to evolve according to the following equations:
    \begin{subequations}
        \begin{eqnarray}
            \frac{d v}{d t} &=& h_v(v,P) + g_{vv}(v,P)\Gamma_1 +
                                          g_{vP}(v,P)\Gamma_2
            \label{power-velocity1}\\
            \frac{d P}{d t} &=& h_P(v,P) + g_{Pv}(v,P)\Gamma_1 +
                                  g_{PP}(v,P)\Gamma_2 .
            \label{power-velocity2}
        \end{eqnarray}
        \label{power-velocity}
    \end{subequations}

In general, the six functions defining vector $\f{h}$ and matrix $\f{g}$ depend on both variables and describe the coupling between each other.
Based on Eqs.~(\ref{KMs}), we can derive both $\f{h}$ and $\f{g}$ from the functions $\f{D}^{(1)}$ and $\f{D}^{(2)}$, which, in turn, are extracted directly from the synthetic data-set by computing the corresponding conditional moments using Eq.~(\ref{DefCoefKM}).
Note that, solving Eq.~(\ref{D2}) for computing the matrix $\f{g}$ yields multiple solutions. If $\f{g}$ is a solution then all matrices of the form $\tilde{\f{g}}=\f{g}\f{O}$ where $\f{O}$ is an orthogonal matrix ($\f{O}\f{O}^T=\f{1}$) are also admissible solutions.

The matrix  $\f{g}$ can therefore be computed as  the ``square root''
of matrix $\f{D}^{(2)}$, i.e.~by diagonalizing $\f{D}^{(2)}$ through a proper 
permutation matrix and ---since all eigenvalues are positive ($\f{D}^{(2)}$ is 
positive  definite)---taking the square root of each eigenvalue and
transforming the matrix back.

Figures \ref{fig4}a-e show the five components of $\f{D}^{(1)}$ and $\f{D}^{(2)}$, i.e. the numeric results for both the drift and the diffusion coefficients computed directly from the generated $P$ and $v$ time series.
The large fluctuations in the region near to maximum power production and wind speed are due to lack of observations. Indeed, the joint PDF for $P$ and $v$ (Fig.~\ref{fig4}f) shows that this region is poorly sampled.

To extract valuable information, next we treat these functions separately. Namely, we consider the drift vector field $(h_v,h_P)$ and the eigenvectors of the diffusion matrix associated to its eigenvalues $\lambda_{max}$ and $\lambda_{min}$. Figure~\ref{fig5} shows the drift vector field in the power production and wind speed state space, restricted to the sampled region defined by the power production curve in Fig.~\ref{fig3}. The solid black line is the performance curve computed from the (P, v) joint probability density function, shown in Fig.~\ref{fig4}f, and defines the most likely power production for a given wind speed. Three different regions can be identified.

Region I in Fig.~\ref{fig5} is characterized by a large wind speed, i.e.~above a ``threshold''
velocity $v_{th}$ that exceeds the rated wind speed of the turbine. Postponing a more detailed description to section \ref{sec:integration}, we define $v_{th}=0.55 v_{max}$, which corresponds to a 97-percentile of the wind speed distribution. For these rare events of high wind speeds, the expected behavior of the wind turbine is to maintain the power production since there is a surplus of energy in the airflow. In this region, the performance curve is roughly constant at $\sim 0.95 P_{max}$. Still, positive power drifts are observed whenever the power production is below the performance curve. The wind speed drift is large in magnitude and always negative, i.e. the drift points towards lower wind velocities.

Region II is characterized by production levels above a transition region of $0.6 P_{max} \lesssim P \lesssim 0.8 P_{max}$, indicated by gray shading in Figs.~\ref{fig5},\ref{fig6}, which is defined in more detail in section \ref{sec:integration} and Fig.~\ref{fig6}. A closer look at Region II enables one to identify a fixed point region ($\f{D}^{(1)}\sim 0$) at high power production levels and 
wind speed $v\sim 0.5v_{max}$,
shown as an encircled area in Fig.~\ref{fig5} near the $v_{th}$. This speed value coincides approximately with the rated wind speed, i.e.~the speed for which the turbine was designed and at which it operates at an optimal regime. It can therefore be concluded from our analysis that the turbine has been well selected, and it remains to be seen if  similar conclusions can be drawn when applying our method to arrays of turbines.

Finally, region III is characterized by frequent low-speed events with a power production below $P \approx 0.6 P_{max}$, containing another attraction point at $v_{mode}  \approx 0.14 v_{max}$.

In previous works\cite{gottschall2007} the drift vector field around
the performance curve is parallel to the power production axis. In 
Fig.~\ref{fig5} the vector field tends to be tilted towards the performance
curve, because the data analyzed was sampled with a much smaller frequency,
and therefore the time between successive measures is sufficiently large
to observe the convergence to the stable fixed points.

Such observations can be more clearly understood by considering Fig.~\ref{fig5} together with the marginal PDF of the wind speed shown in its inset. The distribution of observed values for the wind speed follows 
approximately a Weibull distribution, as is known from the 
literature\cite{weibull}. 
In our case we observe significant deviations at the tail, which shows a
bump. This deviation can be explained with the help of Fig.~\ref{fig2}c
and by recalling that there are correlations between the wind speed with each one of the
main wind directions (as highlighted in section~\ref{sec:data}), resulting in the bimodality of the wind speed.
The bump in the distribution for the wind speed
indicates one of the two modes, namely the one observed at
high wind speeds. 
Therefore, the bimodality of the original data is preserved as bimodality of the Markov chain model.

    The two dominant trends identified in region I are compatible with the expected behavior of the power production control system present on the wind turbine. For high values of wind speed, the controller action upon the blade aerodynamics is capable of sustaining the production level despite the expected decrease of the wind speed. One has to consider, however, the time scales involved. With a 10min resolution of the original data, it is not possible to directly observe the rapid controller action on the blades, only the average behavior of the controller as well as actions that occur on larger time scales, such as the rotation of the tower. However, even at large sampling times, the data set catches some events outside of the power curve and the subsequent conditional moments mirror the controller action that forces the system back on the curve.

    Another important application of our method deals with the diffusion matrix. As explained in the previous section, by diagonalizing the diffusion matrix at each point of the phase space one is able to determine the two eigendirections for diffusion. Being orthogonal to each other, these two eigendirections define an ellipse with major and minor axis proportional to the corresponding eigenvalue. Figure~\ref{fig6} shows the diffusion ellipses in phase space. 
Region I is characterized by the largest ellipses indicating very large fluctuations, while there is an area in region II, which presents small fluctuations and corresponds to the fixed point areas identified in the drift field. In the high-slope region of the power curve, the ellipses degenerate, i.e.~one eigenvalue is negligible when compared to the other ($\lambda_{min}/\lambda_{max}\sim 0$). The inset of Fig.~\ref{fig6} shows in a gray scale the quotient $\lambda_{min}/ \lambda_{max}$ between the smallest and the largest eigenvalue. White corresponds to zero quotient, while values in $[0.2 \dots 1]$ are colored in black. Clearly, a white region indicating a very low ratio of the eigenvalues can be identified, which follows the performance curve shown in Fig.~\ref{fig3}.
Details concerning our error analysis are described
in Append.~\ref{append:errors}.

    \begin{figure}[htb]
        \centering
        \includegraphics[width=0.5\textwidth]{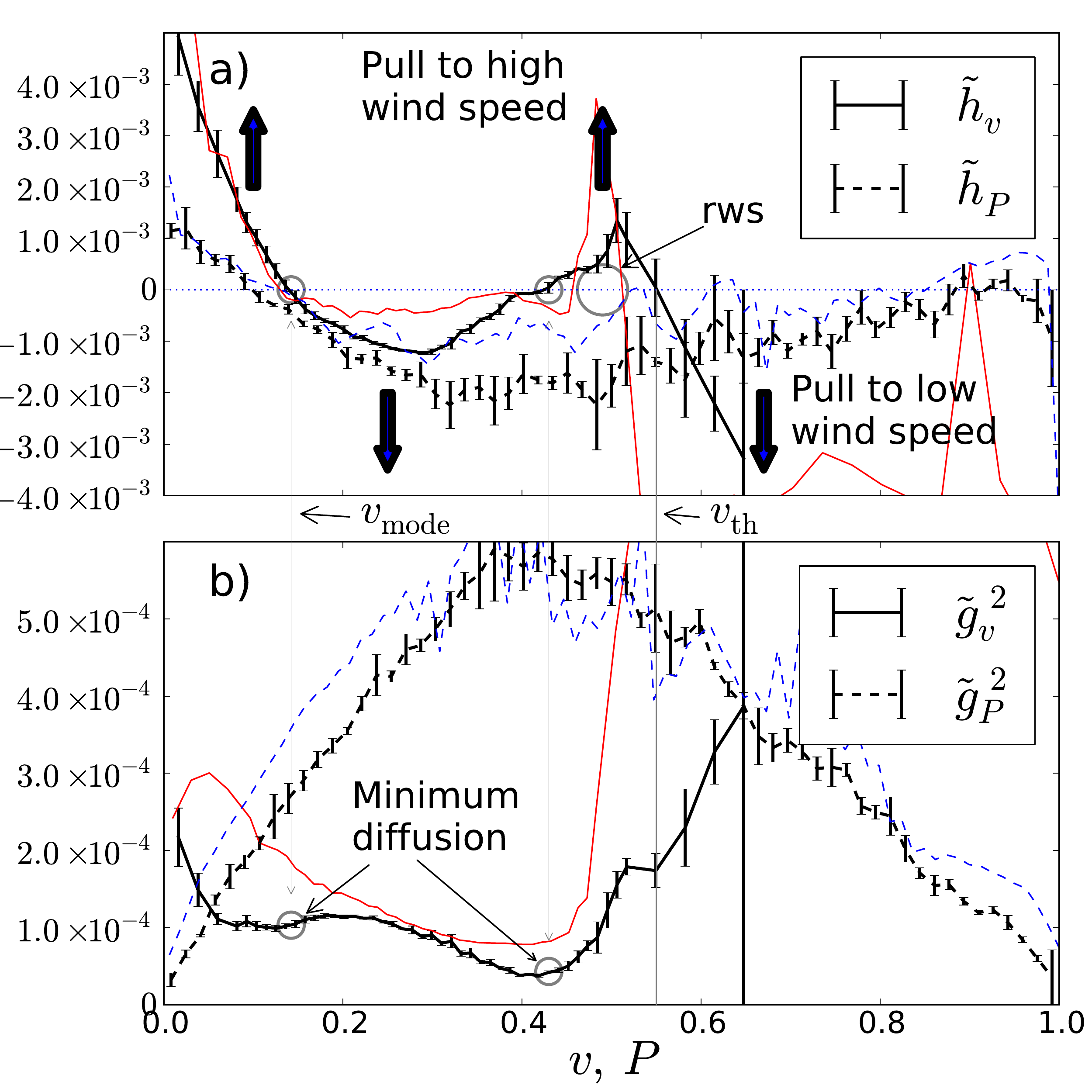}
        \caption{\protect (Color online) Uncovering properties of wind turbines by analyzing
         data series of wind speed $v$ and power $P$ separately (see
         text and Eqs~(\ref{velocityLang}) and (\ref{PowerLang})).
         Horizontal axis indicates the value of $v/v_{max}$ for $\tilde{h}_v$ 
         and $\tilde{g}^2_v$ and the value of $P/P_{max}$ for 
         $\tilde{h}_P$ and $\tilde{g}^2_P$. 
         The large circle denoted as `rws' marks the region of the rated
         wind speed also seen in Figs.~\ref{fig5} and \ref{fig6}.
        Lines without error bars  indicate the 
        same  Kramers-Moyal coefficient functions, this time 
        derived directly from the transition matrix (see text), 
        yielding similar results (red solid line in a): $\tilde{h}_v$, blue dashed line in a): $\tilde{h}_P$, red solid line in b): $\tilde{g}_v^2$, blue dashed line in b): $\tilde{g}_P^2$).   }
        \label{fig7}
    \end{figure}

One remark is appropriate at this point. As we mention above, the procedure described
in this section has been performed on synthetic data generated using a Markov chain model.
One might argue that the drift and diffusion coefficients could be extracted directly
from the measured data. However, as a process in time, the real data also reflects daily and seasonal variations, which hinders and eventually spoils this approach. It has been found (cf.~\ref{sec:performance} ) that the dynamics of the system are reflected by the presence of two maxima in the joint distribution (one at low $v$, near $v_{\mathrm{mode}}$ and one at high $v$, near $v_{\mathrm{th}}$). The two dimensional PDF of the measured data (not shown) closely resembles the one of synthetic data, shown in Fig.~\ref{fig4}f, including  the two maxima. Estimating drift and diffusion coefficients directly from the measured data, however, does not allow to reproduce this distribution (not shown). The MC model has been found to reproduce the relevant statistical  and dynamical features of the fluctuations observed in the real  data\cite{Lopes2012}. Given the facts that {\it i)} the MC model produces a transition matrix that maximizes the likelihood of distributions \cite{Lopes2012},  {\it ii)} its errors are known and small (cf.~\ref{append:errors}),  {\it iii)} it faithfully reproduces the joint PDF, and {\it iv)} the analysis using the MC yields the correct physical dynamics and fixed points (cf.~\ref{sec:performance}), it is reasonable to take the simulated data as the aperiodic process corresponding to the time series of power and wind speed. 

\section{Deriving the performance curve from univariate stochastic dynamics}
\label{sec:integration}

The inset of Fig.~\ref{fig6} shows that along the performance curve one eigenvalue is typically much larger than the other.
This is an indication that in fact $P$ is a function of $v$, which in the case of power 
production and wind speed yields the performance curve drawn in Fig.~\ref{fig3},
\ref{fig5} and \ref{fig6}. 

To see this one first takes $P$ and $v$ as two general variables fulfilling Eqs.~(\ref{power-velocity}) and observes that if 
$P\equiv P(v)$ there are not two independent stochastic forces, but only one, yielding for $v$\cite{gard} and for $P$
  \begin{subequations}
\begin{eqnarray}
        \frac{d v}{d t} &=& {\tilde{h}_v(v)} +
                    \tilde{g}_v(v) \Gamma ,
        \label{velocityLang}\\
        \frac{d P}{d t} &=& {\tilde{h}_P(P)} +
                    \tilde{g}_P(P) \Gamma ,
        \label{PowerLang}
\end{eqnarray}
\label{Lang}
    \end{subequations}
where functions $\tilde{h}$ and $\tilde{g}$ are of course different from the drift and diffusion functions defined 
above in Eqs.~(\ref{power-velocity}), since only one variable is taken into consideration for the stochastic motion 
equation. 

Consequently, the separate analysis of both wind speed and power production enables one to extract valuable insight about the full dynamics and behavior
of the wind turbine with the atmospheric wind.
In fact, the 2D analysis of the performance curve summarized in Figs.~\ref{fig5} and \ref{fig6} can indeed be accessed 
through a one dimensional stochastic analysis of each variable $P$ and $v$ separately.
    \begin{figure}[t!]
        \centering
        \includegraphics[width=0.5\textwidth]{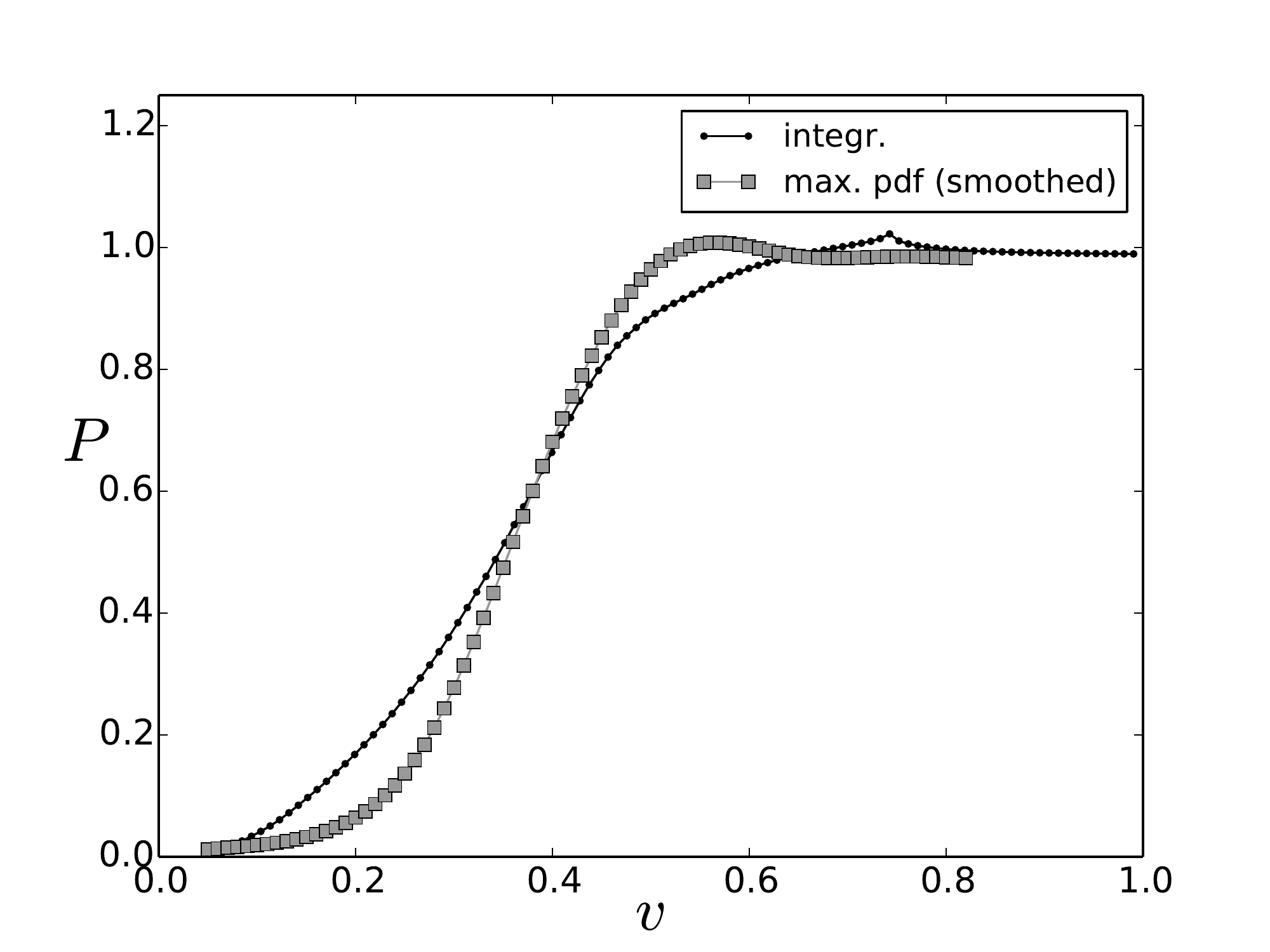}
        \caption{\protect Power production $P$ as function 
           of wind speed $v$ by 
           integration (see Eq.~(\ref{power})). Both Eqs.~(\ref{power-vel})
           are fulfilled, analyzing both series $P$ and $v$ separately
           (see Eqs.~(\ref{velocityLang}) and (\ref{PowerLang})).}
        \label{fig8}
    \end{figure}

Figure \ref{fig7} shows the drift and diffusion of both the wind
speed and power production determined for the model described by 
equations Eqs.~(\ref{velocityLang},\ref{PowerLang}).
The drift of the wind speed $\tilde{h}_v$ has three zeros.
These zeros correspond to three fixed points, two stable ($v\simeq v_{mode}\simeq 0.14 v_{max}$
and $v\simeq v_{th} \simeq 0.55 v_{max}$), and one unstable at $v_{\mathrm{UFP}}\simeq 0.42 v_{max}$.
Thus, for wind speed below $v_ {\mathrm{UFP}}$, the airflow is unstable and unsuited for power production,
while wind speeds above $v_{\mathrm{UFP}}$ promote power production.
The first zero of $\tilde{h}_v$ indicates approximately the mode of the wind speed
 distribution - compare with inset in Fig.~\ref{fig5} - and the other
two zeros mark the transition between two different regions
identified above in Figs.~\ref{fig5} and \ref{fig6}. 
The transition between Region I and Region II is marked by $v_{th}$. The transition between
Region II and III is more subtle and deals with the zero at $v_{\mathrm{UFP}}$ and with the 
functional dependency of the two variables: it is located at the transition region of $0.6 P_{max} \lesssim P \lesssim 0.8 P_{max}$, which corresponds to $v_ {\mathrm{UFP}} \simeq 0.42v_{max}$, cf.~Fig.~\ref{fig5}.
The transition between Region III and Region II is also located at a minimum of the diffusion $\tilde{g}_v$ for the wind velocity. 
Moreover, Fig.~\ref{fig7}a also shows that positive drifts are located at small wind speeds (up to $v_{mode}$) and for Region II (see Figs.~\ref{fig5} and \ref{fig6}). At $v_{\mathrm{UFP}}$, the drift changes to a positive value. In other words, above $v =v_{\mathrm{UFP}}$ the expected change of the wind speed is towards higher values. The drift reaches a maximum on Region II prior to a steep change towards negative values. In this region, the wind speed values are not in the range of the extreme weather conditions and are also not as frequent as the lower wind speed values. However, it is frequent enough to be associated with a commonly repeated pattern, i.e. the bimodal pattern of the wind direction, in which the airflow is mainly induced by thermal differences. This pattern is responsible for  most of the power production in this wind turbine and the main reason for a second attraction point at $v_{th}$. 
For higher wind speed values, the drift changes again to negative values (Region I) since very high wind speed is usually of short duration, i.e. extreme wind gusts.
Moreover, Fig.~\ref{fig7}b shows  where the power stochasticity is maximal, i.e. the region of the highest diffusion values $\tilde{g}_P$, which is the high-slope region of the power curve for $0.3 P_{max} \lesssim P \lesssim 0.6 P_{max}$.

Parallel to our stochastic approach, we also derive coefficients 
$D^{(1)}$ and $D^{(2)}$ directly from the transition matrix. Results are similar
to the ones described above, as one can see in Fig.~\ref{fig7} (thin gray 
solid and dashed lines), where
the deviations from the coefficients derived from our analysis are due to the prescribed binning of phase space, 
i.e.~to the number of states chosen for the transition matrix.

Bringing all the above observations into account, one concludes that there is a strong agreement between the regions defined in the context of Fig.~\ref{fig5} and \ref{fig6} and 
the sign of the wind speed drift.
 
Having analyzed separately both properties, $v$ and $P$, we continue by showing that from the drift and diffusion coefficients,
$\tilde{h}_v$, $\tilde{g}_v$, $\tilde{h}_P$ and $\tilde{g}_P$, one  obtains a functional
dependence between power production and wind speed.
To that end we assume that Eq.~(\ref{velocityLang}) 
holds for $v$ and that the other variable $P$ is an exclusive function 
$P(v)$ of $v$.
Thus, we can take the It\^o-Taylor expansion\cite{gard} of its differential
    \begin{eqnarray}
        dP(v) &=& P(v+dv)-P(v) \cr
        &=& \frac{dP}{dv} dv + \tfrac{1}{2}\frac{d^2P}{dv^2}dv^2 + {\cal O}(dv^3) \cr
        &=& \left ( \frac{dP}{dv}\tilde{h}_v + \tfrac{1}{2}\frac{d^2P}{dv^2}\tilde{g}_v^2
            \right ) dt + \frac{dP}{dv}\tilde{g}_v dw 
    \label{power}
    \end{eqnarray}
    using the differential $dv = {\tilde{h}_v(v)}dt + \tilde{g}_v(v) dw$. Therefore, identifying
    \begin{subequations}
        \begin{eqnarray}
            \tilde{h}_P &=& \frac{dP}{dv}\tilde{h}_v + \tfrac{1}{2}\frac{d^2P}{dv^2}\tilde{g}_v^2
            \label{power-vel1}\\
            \tilde{g}_P &=& \frac{dP}{dv}\tilde{g}_v
            \label{power-vel2}
        \end{eqnarray}
        \label{power-vel}
    \end{subequations}
\begin{widetext}
which can be solved with respect to the two derivatives of $P(v)$ 
yielding the numerical integration scheme as follows:
    \begin{eqnarray}
    P(v+\Delta t)&=&P(v)+\frac{dP}{dv}\vert_{v,P(v)} \Delta v +
                    \frac{1}{2} \frac{d^2P}{dv^2}\vert_{v,P(v)} (\Delta v)^2 + {\cal O}((\Delta v)^3) \cr
&=&P(v)+\frac{}{}
\frac{\tilde{g}_P(P(v))}{\tilde{g}_v(v)} \Delta v +
                    \frac{1}{2} 
\frac{\tilde{h}_P(P(v))\tilde{g}_v(v)-\tilde{h}_v(v)\tilde{g}_P(P(v))}{(\tilde{g}_v(v)^3} (\Delta v)^2 + {\cal O}((\Delta v)^3) .
    \label{Pintegration}
    \end{eqnarray}
\end{widetext}
Figure \ref{fig8} shows the integration of $dP(v)$ for the condition 
$P_0(v_0)=0$ for $v_0=0$. The deviations can be attributed to the fact that 
Eqs.~(\ref{velocityLang}) and (\ref{PowerLang}) are strictly only valid in 
the regions where the eigenvalues of the diffusion matrix show a large 
difference between them, $\lambda_{min} \ll \lambda_{max}$, cf.~inset of 
Fig~\ref{fig5} and \ref{fig6} where $0.2 \leq v \leq v_{th}$. In addition, 
it also holds only on the performance curve, and applying 
Eqs.~(\ref{velocityLang}) and (\ref{PowerLang}) therefore also neglects 
the asymmetry of the drift functions with respect to this curve. Both deviations are a natural consequence of having treated two dependent
variables, $P$ and $v$, as separated stochastic variables. 

However, by doing so, two important features can be  observed. First, the diffusion matrix 
$\mathbf{D}^{(2)}(P,v)$ has rank one, i.e.~one of its eigenvalues can
be neglected in comparison to the other. Second, from the functions $h$ 
and $g$ in Eqs~(\ref{Lang}), it is possible to determine the functional
dependence between both variables. More details are given in 
Append.~\ref{append:dependence}.

\section{Discussion and conclusions}
\label{sec:conclusions}

Investigating a wind turbine from a real wind park, we report the reconstruction of the stochastic performance curve in the variables wind speed and  power production,  using both drift and diffusion coefficients. These coefficients, describing the respective deterministic and stochastic interactions of wind field, turbine aerodynamics, and controller action, are estimated from a synthetic  time series  generated using a Markov Chain model of the original measurement data. We argue that this reconstruction is superior to a direct evaluation of the measurements.

As a main finding, we present the fact that the reconstruction of the power curve using both drift and diffusion coefficients uncovers additional information not visible in an analysis of the drift field alone\cite{anahua2008}, even though we are using measured data of a very low measurement rate as  model input. Specifically, our analysis reveals the existence of various distinct regions in the wind speed--power production plane.

In addition, we have been able to reconstruct the power curve from the drift and diffusion coefficients, using a method which should be able to uncover functional relationships between stochastic variables in a wide range of experimental setups.

It should be noted that it is possible to infer the correlation between $v$ and $P$ from the joint PDF in Fig.~\ref{fig4}f alone; however, such an approach neglects the dynamical behavior of the system. 

Without our approach summarized in Figs.~\ref{fig5}, \ref{fig6} and 
\ref{fig7}, one could not so surely claim the existence of three separated 
regions.
Moreover, attached to these three regions we detected three fixed
points of the dynamics, two of them stable and one unstable. These three 
fixed points are only clearly shown in Fig.~\ref{fig7}, after performing 
the one-dimensional analysis with our method.

In particular, considering the upper stable fixed point, while its coincidence with peak production can be identified directly in the (P, v) joint PDF, 
the analysis sketched in Fig.~\ref{fig7} allows to detect the region 
which belongs to its basin of attraction, within which drift drives 
the trajectories to the upper stable fixed point, whereas out from this 
regions trajectories are pushed to the lower stable fixed point.

Finally, information about how trajectories diffuse in phase space,
i.e. the entire dynamics of the system, can only be obtained completely
after extracting the drift and diffusion fields together with the
fixed points of the drift field and the principal directions of diffusion.
Only after analyzing the separation of diffusion eigenvalues visible in 
Fig.~\ref{fig6} can we  postulate  the existence of a single  diffusive 
force underlying Eqs.~\eqref{velocityLang}-\eqref{partP}.

Although there are limitations in increasing the extracted power of a wind
turbine, typically described by the power coefficient which has a maximum
value given by the Betz limit\cite{windenergbook}, 
our approach may be helpful in obtaining a better understanding of the 
complex dynamics that determines power production in wind turbines.
Indeed, we believe that such a stochastic description if applied to an 
entire wind park, would enable to better quantify the risk associated 
to the estimate of global energy production.
The global energy production of a wind park is usually determined 
by financial constraints, i.e.~by decision making of how much energy  one  must
buy or sell in the market to compensate the energy production fluctuations.
Therefore, we are now extending this methodology in order to consider 
coupled systems of wind turbines in nearby locations, using direct measures 
of power and wind speed as well as simulated data.

\section*{Acknowledgments}

The authors thank Matthias W\"achter, David Kleinhans and
Maria Haase for useful discussions
and GENERG, SA. for providing the original data.
The authors acknowledge partial support under PEst-OE/FIS/UI0618/2011 and 
FCOMP-01-0124-FEDER-016080 and 
FR (SFRH/BPD/65427/2009), TS (SFRH/BD/86934/2012)
and PGL ({\it Ci\^encia 2007}) also thank Funda\c{c}\~ao para a Ci\^encia 
e a Tecnologia (FCT) for financial support.
This work is part of a bilateral cooperation DRI/DAAD/1208/2013 supported by FCT and Deutscher Akademischer Auslandsdienst (DAAD).

\appendix

\section{Estimation of the Markov chain transition matrix}
\label{append:Markovmodel}

The Markov chain transition probabilities were not obtained employing the usual maximum likelihood (ML) estimator, but by using a modified likelihood function which combines information from 1-step and 2-step transitions, as described in this appendix. This modified ML yields lower variance estimates for the transition probabilities (see Appendix \ref{append:errors}).

Let $\{s_i \mid i \in \{0,...,n\}\}$,~$n \in \mathbb{N}$ be the state space 
of the Markov chain model, where each state is a combination of a wind 
power-, speed-, and direction-state. Thus the wind-power, -speed and 
-direction time-series can be transferred into a stream $s$ of states, 
i.e.~$\mathbf{s} = \{
s_{i_0}, s_{i_1}, s_{i_2}, ..., s_{i_{m-2}}, s_{i_{m-1}}, s_{i_{m}}\}$, 
where $m \in \mathbb{N}$ denotes the length of the time-series and 
$s_{i_k}$ denotes the state the Markov process assumes at time $k$ 
with $i_k \in \{1, ...,n\}$, $\forall k \in \{0, ..., m\}$.
Let $p_{i,j}$ denote the probability of the process for moving from 
state $s_i$, at time $k$, to state $s_j$, at time $k+1$.
Then the likelihood function $\mathcal{L}^{(1)}$, i.e.~the probability of the 
observed series of states $\mathbf{s}$, given the transition probabilities 
$p_{i,j}$ for 1-step transitions, is
\begin{equation}
\mathcal{L}^{(1)} = P(s_{i_0}) \cdot p_{i_0, i_1} \cdot p_{i_1, i_2} 
                        \cdots p_{i_{m-2}, i_{m-1}} \cdot p_{i_{m-1}, i_m}.
\label{L}
\end{equation}
Since the probability of the process being in state $s_{i_0}$ is constant, 
the maximum likelihood estimator can thus be written as the maximum
of $\mathcal{L}^{(1)}=\prod_{(i,j) \in \mathcal{S}_1}p_{i,j}$
subjected to $p_{i,j} \geq 0$ and
$\sum_{j = 0}^n{p_{i,j}} = 1$, with $i,j = 1, \ldots, n$,
where $\mathcal{S}_1$ is the set of all 1-step transitions $(i,j)$ observed 
in the state stream $\mathbf{s}$.

Solving the above optimization problem is equivalent to minimizing the 
negative log-likelihood function 
$\mathcal{L}^{(1)}_{log}=-\sum_{(i,j) \in \mathcal{S}_1}\log p_{i,j}$,
subjected to the same constraints. For practical purposes we consider
henceforth the log-likelihood function $\mathcal{L}^{(1)}_{log}$.

Next, we consider only 2-step transitions. Taking $p^{\prime}_{i,j}$ as the probability of the process for moving from
state $s_i$, at time $k$, to state $s_j$, at time $k+2$,
the maximum likelihood estimators are given by the maximum of 
$\mathcal{L}^{(2)}=\prod_{(i,j) \in \mathcal{S}_2}p^{\prime}_{i,j} = 
 \prod_{(i,j) \in \mathcal{S}_2}\sum_{k = 1}^n p_{i,k}p_{k,j}$
with $\mathcal{S}_2=\mathcal{S}^{odd}_2
\cup \mathcal{S}^{even}_2$, 
where
$\mathcal{S}^{odd}_2 = \{(i_1, i_3), (i_3, i_5), ... , (i_{m-3}, i_{m-1})\}$ 
(odd time points) and 
$\mathcal{S}^{even}_2 = \{(i_0, i_2), (i_2, i_4), ... , (i_{m-2}, i_{m})\}$
(even time points), 
assuming without loss of generality that $m$ is even.
Here, 
we use the Chapman-Kolmogorov equations to express the 2-step 
transitions $p^{\prime}_{i,j}$ in terms of 1-step transitions $p_{i,j}$
with the same constraints as above.

The corresponding log-likelihood function, whose minimization defines 
an equivalent optimization problem, is
$\mathcal{L}^{(2)}_{log}=-\sum_{(i,j) \in \mathcal{S}_2}
\log(\sum_{k = 1}^n p_{i,k}p_{k,j})$.

Finally, as described in Ref.~\cite{Lopes2012}, we minimize the objective 
function given by the sum of the two log-likelihood functions above,
namely 
\begin{equation}
\mathcal{L}_{log}=\mathcal{L}^{(1)}_{log}+\mathcal{L}^{(2)}_{log}, 
\label{Lfinal}
\end{equation}
and under the same constraints. 
Higher order Markov models were also tested and not showed considerable
improvement compared to the estimator $\mathcal{L}_{log}$ in 
Eq.~(\ref{Lfinal}).

Notice that the minimization of $\mathcal{L}_{log}$ in Eq.~(\ref{Lfinal}) 
for the series of states $\mathbf{s}$ is equivalent as to minimize 
the 1-step functional $\mathcal{L}^{(1)}_{log}$ of the augmented state 
stream  $\mathbf{s}^{\prime} = 
s_{i_0}, s_{i_1}, s_{i_2}, ..., s_{i_{m-2}}, s_{i_{m-1}}, s_{i_{m}},
S, s_{i_1}, S, s_{i_3}, S, s_{i_5}, ...,$ $s_{i_{m-3}}, S, s_{i_{m-1}}, 
S, s_{i_0}, S, s_{i_2}, S, s_{i_4}, ..., s_{i_{m-2}}, S, s_{i_m}$,
where $S$ stands for voids in the state stream.

Concerning the state space partition,
we tried several different partitions of the 
power-speed-direction-space.  Comparison of the histograms of the original 
data and the data synthesized using these models showed that the model with 
the highest resolution of the binning in the power-speed plane is 
$80 \times 60 \times 12$, which reproduces the original data most accurately.

\section{Error Analysis}
\label{append:errors}

A lower bound for the uncertainty of the transition probabilities is provided by the Cramer-Rao bound (CRB), since there is no simple expression to determine the parameters variance for the 2-step estimator, described in section \ref{append:Markovmodel}. The CRB for maximum likelihood estimators with constraints on the parameter space can be computed by using the following equation \cite{Stoica1998}:
\begin{equation}
        {CRB} = U^T(U H U^T)^{-1}U,
\end{equation}
where $U$ is an orthonormal matrix spanning the null-space of the Jacobian $J$ of the parameters equality constraints and $H$ the Hessian of the objective function, defined in section \ref{append:Markovmodel}.

The uncertainty is linked to the number of observed transitions in the data. Results show that a high value for the CRB ($\sigma_{p_{ij}} > 0.1$) is always associated with infrequent transitions ($<4$ observations, in the 2-yr dataset). For transition probabilities $p_{ij}$ with more than 25 observations, 90\% of the $\sigma_{p_{ij}}$ values are smaller than $0.25  p_{ij}$.

To assess the performance of the 2-step estimator, its CRB is compared with the CRB of the 1-step estimator, applied to the same 2-yr dataset. Results shows that for the transition probabilities with a high CRB value, the 2-step estimator provides a lower bound. In the remaining cases, the difference is not significant.

Another source of errors stems from our use of the direct estimation of the Kramers-Moyal (KM) coefficients from the synthetic time series. This method has been found to introduce three principal types of errors \cite{physrepreview, renner2002, david_diplom}.

The first error accounts for the statistical  variation of counts $N$ in each bin used  for the calculation of the conditional moments. It can be shown \cite{david_diplom} that this error decreases with $1/\sqrt{N}$. Since we can generate synthetic time series of arbitrary length and, therefore, arbitrarily high $N$ in each bin under consideration, we can neglect this error. 

Second, estimation of the KM coefficients uses an expansion of the Fokker-Planck operator in powers of the temporal increment $\tau$, neglecting higher orders in $\tau$. It is known \cite{renner2002} that this finite-time expansion induces an erroneous count $M^{(i)}_E$ of the $i-$th conditional moment $M^{(i)}$, namely $M_{1,E}= \tau D^{(1)} + \frac{\tau^2}{2} ( D^{(1)}D^{(1)}_x + D^{(2)}D^{(1)}_{xx} )$ and 
$M_{2,E}= 2 \tau D^{(2)} + \tau^2( D^{(1)}D^{(1)}   + 2 D^{(2)}D^{(1)}_{x}   + D^{(1)}D^{(2)}_{x} + D^{(2)}D^{(2)}_{xx} )$.


We have calculated these errors numerically, using the estimated KM coefficients $D^{(i)}$ and their numerical first and second derivatives $D^{(i)}_x, D^{(i)}_{xx}$, and found them to be generally within a few percent of the estimated KM coefficients.

A third source of errors is the finite size of the bins used for the calculation of the conditional moments. Again, it can be shown \cite{david_diplom} that this finiteness induces an erroneous count of the moments (and corresponding coefficients)

\begin{equation}
  \label{eq:Error_finite_bin}
  D^{(k)}_E (x_0)  = \frac{\int_{x_0 - \Delta x }^{x_0 + \Delta x }  D^{(k)}(x_0) p(x) \,dx  }{\int_{x_0 - \Delta x }^{x_0 + \Delta x }  p(x) \, dx } \, ,
\end{equation}
 
where $x_0$ and $\Delta x$ are the respective bin centers and bin widths, and $p(x)$ is the stationary  distribution of the stochastic variable. We numerically investigated  this error using fits for both the stationary distribuitions and KM coefficients inside the bins, and found it to be generally  in the few-percent regime, too, increasing considerably only at the edges of the regions investigated, where the KM functions become steeper.

The effect of those two last error sources is indicated by  error bars in Fig.~\ref{fig7}.

\section{Stochastic analysis of two dependent variables}
\label{append:dependence}

In general, whenever the diffusion matrix $\mathbf{D}^{(2)}(P,v)$ 
has rank one, the set of variables $(P,v)$ have in fact only one
independent stochastic source, and therefore Eqs.~(\ref{power-velocity}) 
reduce to Eqs~(\ref{Lang}), where the stochastic force $\Gamma$ is the same for both variables.

Consequently, one can write  the differential of $v$ as
\begin{equation}
dv=\tilde{h}_v dt + \tilde{g}_v dW
\label{dv}
\end{equation}
where the stochastic differential $dW$ is the {\it same} as the one
in Eq.~(\ref{power}). 

Since both variables, $P$ and $v$, are driven by the same stochastic
forces, one can take only one of them as the stochastic variable,
say $v$, and the other one as a function of $v$ and $t$ alone,
$P\equiv P(v,t)$. In that way one incorporates all stochastic 
contributions into $v$.

Mathematically this implies that we can write the differential of
$P$ as
\begin{equation}
dP = \frac{\partial P}{\partial v}dv + \frac{\partial P}{\partial t}dt
\label{power-diff}
\end{equation}
where both partial derivatives are derived from the functions $\tilde{h}$
and $\tilde{g}$ alone (see Eq.~\eqref{Lang}).

Indeed, rewriting Eq.~(\ref{power}) as
\begin{equation}
 dP(v) = \frac{dP}{dv}\left (\tilde{h}_v dt + \tilde{g}_v dW \right ) +
           \tfrac{1}{2}\frac{d^2P}{dv^2}\tilde{g}_v^2 dt
\label{power2}
\end{equation}
and using Eq.~(\ref{dv}) together with the relations in 
(\ref{power-vel}) yields
\begin{equation}
dP = \frac{\tilde{g}_P}{\tilde{g}_v}dv + 
     \left ( 
       \tilde{h}_p-\tilde{h}_v\frac{\tilde{g}_P}{\tilde{g}_v}
     \right ) dt .
\end{equation}

This last equation means that $P$ is a function of $t$ and $v$, which 
contains all stochastic contributions.
Consequently, the partial derivatives in Eq.~(\ref{power-diff}) are
\begin{subequations}
\begin{eqnarray}
\frac{\partial P}{\partial v} &=& 
        \frac{\tilde{g}_P}{\tilde{g}_v}\label{partPpartv}\\
\frac{\partial P}{\partial t} &=&  
       \tilde{h}_p-\tilde{h}_v\frac{\tilde{g}_P}{\tilde{g}_v} .
\end{eqnarray}
\label{partP}
\end{subequations}
The first partial derivative shows that the variation of the production 
power by speed variations  equals the quotient of the corresponding
diffusion amplitudes in time.
The second partial derivative describes the local power production which
is given by the power production drift $\tilde{h}_p$, after subtracting 
the contribution of the wind speed drift $\tilde{h}_v$ in the variation 
of $P$ due to $v$.
        


\end{document}